%% file: conference_101719.tex
\documentclass[conference]{IEEEtran}
\IEEEoverridecommandlockouts
\IEEEpubid{\makebox[\columnwidth]{© 2026 NVIDIA. All rights reserved. \hfill}
\hspace{\columnsep}\makebox[\columnwidth]{ }}
\usepackage[nolist]{acronym} 
\usepackage{cite}
\usepackage{comment}
\usepackage{amsmath,amssymb,amsfonts}
\usepackage{algorithmic}
\usepackage{graphicx}
\usepackage{comment}
\usepackage{textcomp}
\usepackage{xcolor}
\usepackage{floatrow} 
\usepackage{wrapfig}
\usepackage{color}
\usepackage{verbatim}
\usepackage{adjustbox}
\usepackage{hyperref}
\usepackage{tabularx}
\usepackage{tikz}
\usepackage{pifont}
\usetikzlibrary{positioning,shapes.geometric,fit}

\usepackage{url}

\usepackage{subfigure}
\usepackage{booktabs}
\usepackage{multirow}
\usepackage{textgreek}
\usepackage{enumitem}
\usepackage[perpage]{footmisc}

\usepackage{listings}
\usepackage[colorinlistoftodos]{todonotes}

\def\BibTeX{{\rm B\kern-.05em{\sc i\kern-.025em b}\kern-.08em
    T\kern-.1667em\lower.7ex\hbox{E}\kern-.125emX}}

\newcommand{\code}[1]{{\small\texttt{#1}}}

\input{macros.tex}
\input{acros.v2.tex}

\begin{document}

\pagestyle{plain}  

\title{NCCL EP: Towards a Unified Expert Parallel Communication API for NCCL\thanks{NCCL GitHub Repository: \url{https://github.com/NVIDIA/nccl}}}

\input{authors}

\maketitle

\input{text/0-abstract}

\input{text/1-introduction}

\input{text/2-background}

\input{text/3-nccl-ep}

\input{text/4-ll-kernels}

\input{text/5-ht-kernels}

\input{text/6-integration}

\input{text/7-evaluation}

\input{text/8-related-work}

\input{text/9-conclusions}

\input{text/ack}

\bibliographystyle{IEEEtran}

\bibliography{references/ncclep.bib}

\end{document}

%% file: macros.tex
\newcommand{\openshmem}[1][]{%
 {Open\-SHMEM\ifthenelse{\equal{#1}{}}{}{~#1}}}

%% file: acros.v2.tex
\begin{acronym}
\acro{PVM}{Parallel Virutal Machine}
\acro{PICL}{Portable Instrumented Communication Library}
\acro{SSID}{Synchronization Sequence Identification}
\acro{phit}{physical units}
\acro{BTE}{Byte Transfer Engine}
\acro{FMA}{Fast Memory Access}
\acro{MPI}{\emph{Message Passing Interface}}
\acro{UPC}{\emph{Unified Parallel C}}
\acro{PGAS}{Partitioned Global Address Space}
\acro{SIMD}{\emph{Single Instruction Multiple Data}}
\acro{MIMD}{\emph{Multiple Instruction Multiple Data}}
\acro{HPC}{\emph{High Performance Computing}}
\acro{API}{\emph{Application Programming Interface}}
\acro{RDMA}{\emph{Remote Direct Memory Access}}
\acro{AMO}{\emph{Atomic Memory Operations}}
\acro{CRC}{\emph{Cyclic Redundancy Check}}
\acro{MDH}{\emph{Memory Descriptor Handle}}
\acro{PE}{\emph{Programming Element}}
\acro{ORNL}{Oak Ridge National Laboratory}
\acro{UNM}{University of New Mexico}
\acro{MPI}{Message Passing Interface}
\acro{PGAS}{Partioned Global Address Space}
\acro{HPC}{High Performance Computing}
\acro{UCCS}{Universal Common Communication Substrate}
\acro{PROMPI}{Protocol Reconfiguration and Optimization system for \ac{MPI}}
\acro{LANL}{Los Alamos National Laboratory}
\acro{DOE}{Department of Energy}
\acro{DoD}{Department of Defense}
\acro{OLCF}{\emph{Oak Ridge Leadership Computing Facility}}
\acro{QMC}{\emph{Quantum Monte Carlo}}
\acro{CPU}{\emph{Central Processing Unit}}
\acro{SpT}{QMC Samples per Thread}
\acro{AMR}{\emph{Adaptive Mesh Refinement}}
\acro{WL-LSMS}{\emph{Wang Landau - Locally Self-consistent Multiple Scattering}}
\acro{WL}{\emph{Wang Landau}}
\acro{LSMS}{\emph{Locally Self-consistent Multiple Scattering}}
\acro{rLIZ}{\emph{Local Interaction Zone Radius }}
\acro{LAMMPS}{\emph{Large-scale Atomic/Molecular Massively Parallel Simulator}}
\acro{NRDF}{\emph{Non-equilibrium Radiation Diffusion}}
\acro{CESM}{\emph{Community Earth System Model}}
\acro{Vampir}{\emph{Vampir Tool Suite}}
\acro{SAMRAI}{\emph{Structured Adaptive Mesh Refinement Application Infrastructure}}
\acro{CHARMM}{Chemistry at HARvard Macromolecular Mechanics}
\acro{PPPM}{particle-particle particle-mesh}
\acro{IOFSL}{I/O Forwarding Scalability Layer}
\acro{PE}{Processing Element}
\acro{APP}{Application Time}
\acro{COMP}{Computation Time}
\acro{COL}{Collective}
\acro{GPU}{Graphical Processing Unit}
\acro{SIMD}{\emph{Single Instruction Multiple Data}}
\acro{MIMD}{\emph{Multiple Instruction Multiple Data}}
\acro{API}{\emph{Application Programming Interface}}
\acro{RDMA}{\emph{Remote Direct Memory Access}}
\acro{UPC}{\emph{Unified Parallel C}}
\acro{SPMD}{\emph{Single Program Multiple Data}}
\acro{MTTI}{Mean Time Between Interrupts}
\acro{CAF}{Co-Array Fortran}
\acro{SHMEM}{Symmetric Hierarchical Memory}
\acro{FLOPS}{FLoating-point Operations Per Second} 
\acro{MPMD}{Multiple Program Multiple Data} 
\acro{SM}{Shared-Memory} 
\acro{OS}{Operating System} 
\acro{UH}{University of Houston} 
\acro{SNL}{Sandia National Laboratory} 
\acro{ANL}{Argonne National Laboratory} 
\acro{NOP}{No Operation} 
\acro{NUMA}{Non Uniform Memory Access}
\acro{BTL}{Byte Transfer Layer}
\acro{PML}{Point-to-point Management Layer}
\acro{BML}{BTL Management Layer}
\acro{MCA}{Modular Component Architecture}
\acro{DAG}{\emph{Directed Acyclic Graph}}
\acro{BCOL}{\emph{Basic Collective}}
\acro{ML}{\emph{Messaging Layer}}
\acro{P2P}{\emph{Point-to-Point}}
\acro{SM}{\emph{Shared-Memory}}
\acro{MB}{\emph{Megabyte}}
\acro{ORTE}{Open Run-time Environment}
\acro{OPAL}{Open Portable Access Layer}
\acro{IBM}{International Business Machines Corportation}
\acro{HP}{Hewlett-Packard}
\acro{SGI}{Silicon Graphics, Inc.}
\acro{DMAPP}{Distributed Memory Applications}
\acro{VM}{Virtual Machine}
\acro{TCO}{Total Cost of Ownership}
\acro{EC2}{Elastic Compute Cloud}
\acro{NERSC}{National Energy Research Scientific Computing Center}
\acro{ALCF}{Argonne Leadership Computing Facility}
\acro{MR}{\emph{MapReduce}}
\acro{NIST}{National Institute of Standards and Technology}
\acro{10GbE}{10-gigabit Ethernet}
\acro{CUDA}{\emph{Compute Unified Device Architecture}}
\acro{SMs}{Streaming Multiprocessors}
\acro{FFT}{Fast Fourier transform}
\acro{UCX}{Unified Communication X}
\acro{UCS}{UC-Services}
\acro{UCT}{UC-Transports}
\acro{UCP}{UC-Protocols}
\acro{TL}{Transport Layer}
\acro{RMA}{Remote Memory Access}
\acro{PMI}{Process Manager Interface}
\acro{SMSG}{Short message}
\acro{BTE}{Block Transfer Engine}
\acro{SSCA}{Scalable Synthetic Compact Applications}
\acro{RTE}{Run Time Environment}
\acro{SharP}{SHARed data-structure centric Programming abstraction}
\acro{HBM}{High-bandwidth Memory}
\acro{NVRAM}{non-volatile random access memory}
\acro{LLNL}{Lawerence Livermore National Laboratory}
\acro{MD}{\emph{Memory Domains}}
\acro{DT}{\emph{Data Tiers}}
\acro{UMA}{\emph{Unified Memory Allocator}}
\acro{UMI}{\emph{Unified Memory Interface}}
\acro{KV}{\emph{Key-Value}}
\acro{C/R}{\emph{Checkpoint/Restart}}
\acro{ADIOS}{Adaptable IO System}
\acro{MDS}{Metadata Server}
\acro{TEPS}{Traversed Edges Per Second}
\acro{UCX}{Unified Communication X}
\acro{UCS}{UC-Services}
\acro{UCT}{UC-Transports}
\acro{UCP}{UC-Protocols}
\acro{TL}{Transport Layer}
\acro{TLS}{Thread Local Storage}
\acro{SSCA}{Scalable Synthetic Compact Applications}
\acro{SHARP}{Mellanox Scalable Hierarchical Aggregation and Reduction Protocol}
\acro{NCCL}{NVIDIA Collective Communication Library}
\acro{RCCL}{ROCm Communication Collectives Library}
\acro{ANL}{Argonne National Laboratory}
\acro{LANL}{Los Alamos National Laboratory}
\acro{ORNL}{Oak Ridge National Laboratory}
\acro{SNL}{Sandia National Laboratory}

\acro{HCOLL}{Hierarchical Collectives}
\acro{HCA}{Host Channel Adapter}
\acro{RMA}{Remote Memory Access}
\acro{MEMIC}{Memory Mapped to InterConnect}
\acro{BAR}{Base Address Registers}
\acro{UCC}{Unified Collective Communication}
\acro{DL}{Deep Learning}
\acro{EE}{Execution Engine}
\acro{CL}{Collective Layer}
\acro{TL}{Team Layer}
\acro{DPU}{Data Processing Unit}
\end{acronym}

%% file: authors.tex
    
\author{
\IEEEauthorblockN{
    Amos Goldman,
    Nimrod Boker,
    Maayan Sheraizin,
    Nimrod Admoni,
    Artem Polyakov,\\
    Subhadeep Bhattacharya,
    Fan Yu,
    Kai Sun,
    Georgios Theodorakis,
    Hsin-Chun Yin,
    Peter-Jan Gootzen,
    Aamir Shafi,\\
    Assaf Ravid,
    Salvatore Di Girolamo,
    James Dinan,
    Xiaofan Li,
    Manjunath Gorentla Venkata,
    Gil Bloch
}
\vspace{0.4em}
\IEEEauthorblockA{
    NVIDIA Corporation\\
    \textit{\small\{amgoldman, nboker, msheraizin, nadmoni, artemp, subhadeepb, fayu, kais,}\\
    \textit{\small gtheodorakis, hsinchuny, pgootzen, ashafi, aravid, sdigirolamo, jdinan, xiaofanl, manjunath, gil\}@nvidia.com}
}
}

%% file: text/0-abstract.tex
\begin{abstract}%

Mixture-of-Experts (MoE) architectures have become essential for scaling large
language models, driving the development of specialized device-initiated
communication libraries such as DeepEP, Hybrid-EP, and others. These libraries
demonstrate the performance benefits of GPU-initiated RDMA for MoE dispatch and
combine operations.

This paper presents \textbf{NCCL EP} (Expert Parallelism), a ground-up MoE
communication library built entirely on NCCL's Device API. NCCL EP provides
unified \texttt{ncclEpDispatch} and \texttt{ncclEpCombine} primitives with both
C and Python interfaces, supporting Low-Latency (LL) mode for inference decoding
and High-Throughput (HT) mode for training and inference prefill. LL targets small
batch sizes (1--128 tokens) using direct all-to-all RDMA+NVLink mesh connectivity with
double-buffered communication for overlapping dispatch and combine phases. HT
targets large batches (4096+ tokens) using hierarchical communication that
aggregates tokens within NVLink domains before inter-node RDMA transmission.
Both modes leverage Device API for both intra- and inter-node communications,
taking advantage of its topology awareness and optimized GPU-initiated implementation.

We evaluate NCCL EP on an H100-based cluster across multi-node
configurations, demonstrating competitive LL kernel performance and presenting
end-to-end results with vLLM integration. By building MoE communication natively
within NCCL, NCCL EP provides a supported path for expert parallelism on
current and emerging NVIDIA platforms.

\end{abstract}%


%% file: text/1-introduction.tex
\section{Introduction}
\label{sec:introduction}

Mixture-of-Experts (MoE) architectures have become essential for scaling Large Language Models (LLMs) beyond the computational limits of dense 
transformers~\cite{shazeer2017outrageously,lepikhin2021gshard}. 
By individually routing tokens to specialized subsets of experts,
MoE models like DeepSeek-V3~\cite{deepseekai2024deepseekv3} and 
Mixtral~\cite{jiang2024mixtral} achieve superior parameter efficiency and 
computational capacity. However, expert parallelism (EP) introduces communication 
patterns fundamentally different from conventional collectives: fine-grained, 
dynamic all-to-all exchanges where each token is routed to a unique subset of experts,
selected based on learned gating decisions. 
The corresponding \emph{dispatch} (sending tokens to 
experts) and \emph{combine} (gathering results back) operations account for a 
significant fraction of end-to-end latency and require specialized 
optimization~\cite{rajbhandari2022deepspeed,hwang2023tutel}.

One common approach to MoE communication uses NCCL's AllToAll collective 
within a CPU-orchestrated pipeline: tokens are permuted according to routing 
decisions, communicated via AllToAll, and unpermuted after expert computation. 
This \emph{AllToAll dispatcher} pattern, supported in frameworks like 
Megatron-Core~\cite{megatron-core} and Tutel~\cite{hwang2023tutel}, integrates 
naturally with existing communication collectives but adds 
overhead compared to fused device-initiated approaches.

Recent work on \emph{device-initiated} MoE 
communication~\cite{deepep2025,pplx-kernels,hybridep} demonstrated that fusing 
these discrete steps into unified dispatch and combine kernels---where GPUs 
invoke RDMA primitives directly without returning control to the host---can 
significantly reduce latency by enabling in-kernel quantization, overlapped 
data movement, and optimized output layouts.

Despite these advances, existing device-initiated MoE libraries 
face deployment challenges due to ecosystem fragmentation and API complexity. 
DeepEP uses NVSHMEM~\cite{nvshmem} with direct InfiniBand Global Data Access 
(IBGDA) primitives, exposing separate Python APIs for low-latency and 
high-throughput modes.
DeepEP derivatives (i.e. Perplexity’s pplx-kernels~\cite{pplx-kernels})
introduce their custom modifications of the MoE interfaces,
incompatible with each other.
While these libraries achieve exceptional performance 
through fine-grained RDMA control, they require separate communication 
stacks distinct from NCCL---the de facto standard for GPU collective 
communication---and expose mode-specific APIs that complicate application 
development. This fragmentation forces customers to maintain multiple 
runtime environments, learn different interfaces for each mode, and forgo 
NCCL's topology awareness, elasticity, and fault-tolerance mechanisms.

\begin{figure}[t]
\centering
\includegraphics[width=0.85\columnwidth]{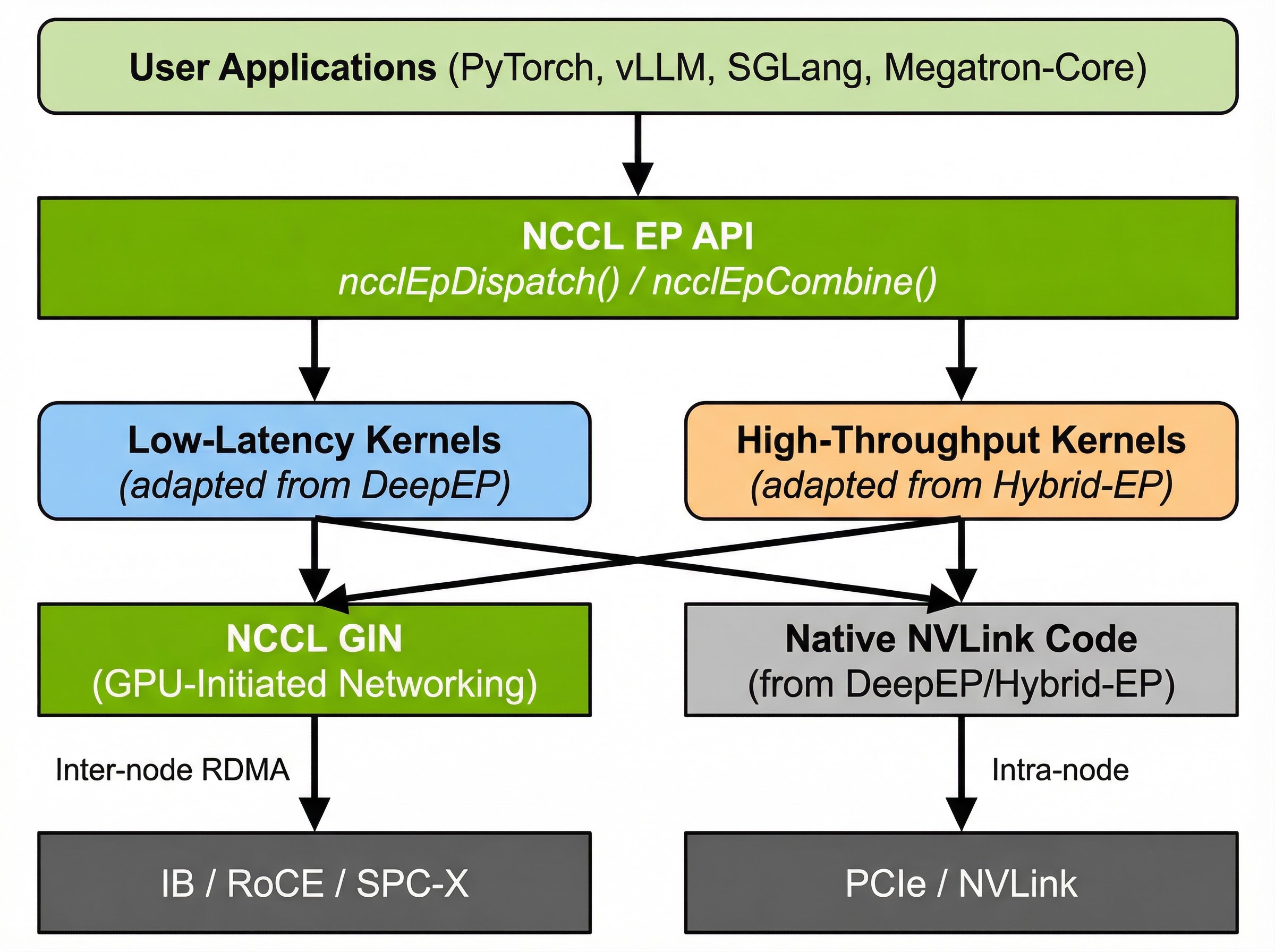}
\caption{NCCL EP architecture: low-latency kernels adapted from 
DeepEP~\cite{deepep2025} and high-throughput kernels adapted from 
Hybrid-EP~\cite{hybridep} use NCCL GIN (GPU-Initiated Networking) for 
inter-node RDMA, while retaining their native NVLink implementations 
for intra-node communication.}
\label{fig:nccl-ep-arch}
\end{figure}

MoE workloads have different performance requirements depending on the use 
case. During \emph{training}, large batches of tokens (4096+) flow through the 
network, making bandwidth utilization the primary concern. During 
\emph{inference}, the workload splits into two phases: \emph{prefilling} 
processes the input prompt with many tokens in parallel (similar to 
training), while \emph{decoding} generates tokens autoregressively one at a 
time per sequence, making latency the critical metric.
In this scenario, the batch sizes are small (1--128 tokens per GPU)
and the latency becomes the primary focus.

This paper presents \textbf{NCCL EP}, a ground-up MoE 
communication library built entirely on NCCL's Device 
API~\cite{nccl2.28,ncclgin}.
Inspired by modern device-initiated MoE libraries, 
NCCL EP brings their performance benefits into the NCCL ecosystem with a 
cleaner interface. Unlike existing solutions that expose separate APIs for 
different operational modes, NCCL EP provides \emph{unified} 
\texttt{ncclEpDispatch} and \texttt{ncclEpCombine} primitives that 
transparently select the appropriate algorithm based on workload 
characteristics. The library offers both C and Python interfaces, providing 
flexibility for framework integration while abstracting hardware complexity. 

Internally, NCCL EP adapts proven kernel designs replacing their 
communication backend with NCCL Device API (Figure~\ref{fig:nccl-ep-arch}).
To support existing MoE configurations, it provides two algorithm modes, 
selected at group creation time:

\noindent\textbf{Low-latency (LL) Mode} for inference decoding.
The design is derived from DeepEP~\cite{deepep2025} low-latency kernel
with memory footprint optimizations.
The implementation employs full all-to-all
mesh connectivity with  double-buffered communication.

\noindent\textbf{High-throughput (HT) Mode} for training and inference 
prefilling. The HT implementation uses hierarchical communication where 
intra-node NVLink aggregates tokens before inter-node RDMA transmission. 
The kernel design is adapted from Hybrid-EP~\cite{hybridep}.

\subsection{Contributions}
\label{sec:introduction:contributions}

Our main contribution is the NCCL EP design and implementation that
addresses both API usability and kernel-level performance through
the following \emph{key design elements}.

\begin{enumerate}[label=\roman*)]

    \item\textbf{Unified API with Algorithm Modes.} NCCL EP provides a 
    unified \texttt{ncclEpDispatch}/\texttt{ncclEpCombine} interface for 
    both training and inference applications. The algorithm mode 
    is specified or (in the future) automatically detected
    at group creation time (\texttt{ncclEpCreateGroup}), 
    allowing applications to switch modes without code changes.
    
    \item\textbf{Two-Tier Resource Management.} NCCL EP separates 
    long-lived resources from per-operation state through a group/handle 
    architecture. The \texttt{ncclEpGroup\_t} manages communication buffers 
    and network connections shared across operations, while 
    \texttt{ncclEpHandle\_t} captures per-forward-pass routing decisions. 
    This design amortizes setup costs while accommodating the dynamic 
    routing patterns inherent to MoE architectures.
    
    \item\textbf{Device-Initiated Communication.} Both LL and HT 
    rely on GPU-initiated communication implemented via
    NCCL Device API. Excluding the CPU from the data path enables
    asynchronous RDMA communication.
    
    \item\textbf{Hybrid Communication Paths.} NCCL EP automatically 
    selects optimal transport: NVLink for intra-node 
    communication and RDMA for inter-node, with graceful fallback to 
    PCIe when NVLink is unavailable.

    \item\textbf{Optimized LL design.} We propose a memory-optimized
    version of the LL kernels that achieves an order-of-magnitude
    reduction in memory footprint.

\end{enumerate}

To evaluate NCCL EP in a production inference serving system,
we provide its \emph{integration} with vLLM and Megatron-LM frameworks.
Building on top of that,
we conduct \emph{comprehensive evaluation} against 
DeepEP across 1--8 node configurations on an H100-based cluster, 
demonstrating competitive LL kernel performance and presenting 
end-to-end results with vLLM integration.

The remainder of this paper is organized as follows.
Section~\ref{sec:background} provides background on MoE architectures, 
GPU communication, and the NCCL Device API.
Section~\ref{sec:nccl-ep} presents NCCL EP's architecture and API design.
Section~\ref{sec:ll-kernels} details the low-latency kernel implementation.
Section~\ref{sec:ht-kernels} describes the high-throughput kernel design.
Section~\ref{sec:integration} discusses integration with ML frameworks 
(vLLM, Megatron-LM) and deployment considerations.
Section~\ref{sec:results} evaluates NCCL EP's performance against DeepEP 
across multi-node GPU clusters.
Section~\ref{sec:related} reviews related work in MoE communication and 
GPU-initiated networking.
Section~\ref{sec:conclusion} concludes with lessons learned and future 
directions.

%% file: text/2-background.tex
\section{Background}
\label{sec:background}

This section provides the foundational concepts required to understand 
NCCL EP's design: MoE communication patterns, the NCCL Device API that 
serves as our implementation foundation, and the current landscape of 
specialized MoE communication libraries.

\subsection{Transformer architecture}
\label{sec:background:transformers}

Modern LLMs are based on the Transformer architecture~\cite{transformers}. A Transformer processes a sequence of tokens using a stack of layers composed of self-attention and feed-forward network (FFN) blocks. Self-attention captures long-range dependencies and contextual relationships between tokens. The FFN applies nonlinear transformations independently to each token representation, increasing the expressive capacity of the model. 

During the \emph{forward pass} (training and inference), token embeddings propagate through the Transformer layers to produce output logits representing predicted token probabilities. During the \emph{backward pass} (in training), the gradients of the training loss are computed via backpropagation and used to update the model parameters. Repeating this process over large datasets allows LLMs to learn statistical patterns of natural language.

\subsection{MoE Communication Patterns}
\label{sec:background:moe}

The MoE models extend the Transformer architecture,
implementing the FFN as a network of experts,
distributed among available 
GPUs~\cite{shazeer2017outrageously,lepikhin2021gshard}.
During forward 
propagation, token embeddings are routed to a subset of assigned experts, and expert responses are gathered back.
On the backward pass, the loss gradients are routed to the experts, who provided answers, to update their parameters.
In addition, the MoE router parameters are updated.

\textbf{Dispatch} routes tokens from source GPUs to the 
GPUs hosting their assigned experts. A gating network selects top-$k$ experts 
per token, producing routing indices that define a dynamic, irregular communication 
pattern. Each GPU sends a unique subset of tokens to each destination.

\textbf{Combine} aggregates the expert outputs back to the original 
token locations. When tokens visit multiple experts 
(top-$k > 1$), combine performs a weighted summation of expert outputs 
according to gating weights.

These operations differ fundamentally from standard collective 
communication. Unlike AllReduce or AllGather where all participants 
exchange the same data volumes, MoE dispatch and combine exhibit:
\begin{itemize}
    \item \textbf{Dynamic routing}: Communication patterns change every 
        forward pass based on learned gating decisions.
    \item \textbf{Irregular message sizes}: Each GPU pair exchanges
        different amounts of data depending on routing.
    \item \textbf{Load imbalance}: Popular ``hot'' experts receive more 
        tokens, creating asymmetric communication volumes.
    \item \textbf{MoE-specific compute component}: Fusion requirements like dispatch quantizing tokens and combine doing custom reduce.
\end{itemize}

Standard NCCL AllToAll treats all ranks uniformly and cannot efficiently 
handle these irregular patterns, motivating specialized MoE communication 
libraries.

\subsection{NCCL Device API}
\label{sec:background:deviceapi}

NCCL 2.28 introduces the Device API~\cite{nccl2.28,ncclgin}, 
enabling GPU kernels to initiate communication directly
without CPU involvement.
The Device API comprises three modules:

\noindent\textbf{Load/Store Accessible (LSA)} provides direct memory 
access to remote GPU memory over NVLink. Applications can identify
that a peer GPU memory is locally accessible and obtain the corresponding pointer
through \texttt{ncclCommGetLocalSymm}.
Such a pointer can be subsequently used in standard CUDA load/store operations.
LSA delivers the full NVLink bandwidth (900~GB/s bidirectional on H100) with minimal latency, 
making it ideal for intra-node token transfers.

\noindent\textbf{GPU-Initiated Networking (GIN)} enables GPU kernels to 
perform RDMA operations to remote nodes. GIN exposes device-callable 
primitives including:
\begin{itemize}
    \item \texttt{ncclGin\_Put}: Write data to remote GPU memory
    \item \texttt{ncclGin\_Get}: Read data from remote GPU memory  
    \item \texttt{ncclGin\_SignalAdd}: Atomically increment a remote 
        signal for synchronization
    \item \texttt{ncclGin\_WaitOnSignal}: Block until a signal reaches 
        a threshold
\end{itemize}

GIN operations execute entirely on the GPU, eliminating CPU proxy thread 
overhead and enabling fine-grained interleaving of computation and 
communication within a single kernel. This is essential for MoE workloads 
where per-token communication decisions benefit from device-side control.

\noindent\textbf{Multimem} provides collective load/store operations 
across multiple GPUs, useful for broadcast and reduction patterns within 
NVLink domains.

By building on NCCL's Device API rather than standalone libraries like 
NVSHMEM or DOCA GPUNetIO, NCCL EP inherits NCCL's topology detection, 
network plugin architecture, and ecosystem compatibility while gaining 
device-initiated communication capabilities.

\subsection{The MoE Communication Landscape}
\label{sec:background:landscape}

The demand for efficient MoE communication has produced a diverse 
ecosystem of specialized libraries, each optimized for specific hardware 
configurations or use cases (also see Section~\ref{sec:related}).
Table~\ref{tab:moe-libraries} compares 
key features across major solutions.

\begin{table*}[t]
\centering
\caption{MoE Communication Libraries: Feature Comparison}
\label{tab:moe-libraries}
\footnotesize
\begin{tabular}{@{}lccccc p{2.4cm} p{3.cm}@{}}
\toprule
\textbf{Library} & \textbf{Unified API} & \textbf{LL Mode} & \textbf{HT Mode} & \textbf{C API} & \textbf{Python} & \textbf{Inter-node} & \textbf{Target Workload} \\
\midrule
DeepEP~\cite{deepep2025} & No (separate) & Yes & Yes & No & Yes & NVSHMEM/IBGDA & Training + Inference \\
Hybrid-EP~\cite{hybridep} & No (HT only) & No & Yes & Yes & Yes & IBGDA & Training (Megatron) \\
pplx-kernels~\cite{pplx-kernels} & No (separate) & Yes & Yes & Rust/C++ & Yes & RDMA (CX7, EFA) & Inference \\
UCCL-EP~\cite{uccl-ep} & No (separate) & Yes & Yes & Yes & Yes & CPU proxy & Multi-vendor portable \\
NCCLX~\cite{ncclx} & Yes & Yes & Yes & Yes & Yes & CTran (IB/RoCE) & Meta-scale~(100k+ GPUs) \\
MORI~\cite{mori} & Modular & Yes & Yes & Yes & Yes & rocSHMEM & AMD GPUs \\
\midrule
\textbf{NCCL EP} & \textbf{Yes} & \textbf{Yes} & \textbf{Yes} & \textbf{Yes} & \textbf{Yes} & \textbf{NCCL GIN} & \textbf{Training + Inference} \\
\bottomrule
\end{tabular}
\end{table*}

These libraries differ in API style, transport mechanisms, and 
target workloads, and no single solution covers all use cases within 
one framework. This diversity introduces several considerations for 
production deployments:
\begin{itemize}
    \item \textbf{Ecosystem isolation}: Most libraries operate outside 
        NCCL, requiring separate initialization, memory management, and 
        debugging infrastructure
    \item \textbf{Hardware coupling}: Some solutions require specific 
        transport backends (e.g., NVSHMEM, IBGDA), tying deployments to 
        particular NIC configurations
    \item \textbf{API complexity}: Libraries expose separate interfaces 
        for low-latency and high-throughput modes, complicating 
        framework integration
\end{itemize}

In contrast, NCCL EP provides a unified MoE 
communication API built entirely on the standard NCCL ecosystem.

%% file: text/3-nccl-ep.tex
\section{NCCL EP: Design and API}
\label{sec:nccl-ep}

This section presents the design and API of NCCL EP.
We describe the resource management model, tensor abstraction,
core operations, and algorithm modes 
that enable efficient dispatch and combine operations
for MoE workloads.

\subsection{Design Philosophy}
\label{sec:nccl-ep:design}

NCCL EP design reflects several key principles drawn from NCCL programming model:

\noindent\textbf{Native C API with Python Bindings.}
Unlike existing MoE libraries that expose Python/PyTorch interfaces over C++ internals, 
NCCL EP offers a first-class C API aligned with NCCL syntax and conventions. 
This supports direct integration into C++ runtimes (e.g., TensorRT-LLM, native vLLM backends) 
while still enabling Python workflows via a ctypes-based wrapper.
The C API ensures that NCCL EP can serve as a foundational library for 
framework developers without requiring PyTorch dependencies.

\noindent\textbf{User-Controlled Resource Lifecycle.}
Following the NCCL design, users explicitly control when resources are 
allocated and destroyed. All resources are created explicitly
and persist until destroyed. This model provides 
predictable memory usage and enables integration with custom memory 
allocators, which are critical for inference serving systems that manage 
memory pools across multiple models.

\noindent\textbf{Stream-Based Asynchronous Execution.}
All NCCL EP operations execute asynchronously on CUDA streams, consistent 
with NCCL execution model, 
avoiding implicit synchronization points that harm performance.

\noindent\textbf{Hardware Abstraction.}
NCCL EP abstracts both the communication and computation components
of dispatch and combine operations,
enabling seamless support for future features (like extended NVLink domains)
or alternative network backends.

\noindent\textbf{GPU-Accelerated Communication.}
NCCL EP takes advantage of massive parallelism
offered by NVIDIA GPUs accelerating both
communication and computation components in a form of
fused CUDA kernels.
The kernels are mapped into GPU architecture,
distributing token processing between available
Streaming Multiprocessors (SMs).

\subsection{Core Operations}
\label{sec:nccl-ep:operations}

\begin{figure}[t]
\centering
\includegraphics[width=0.95\columnwidth]{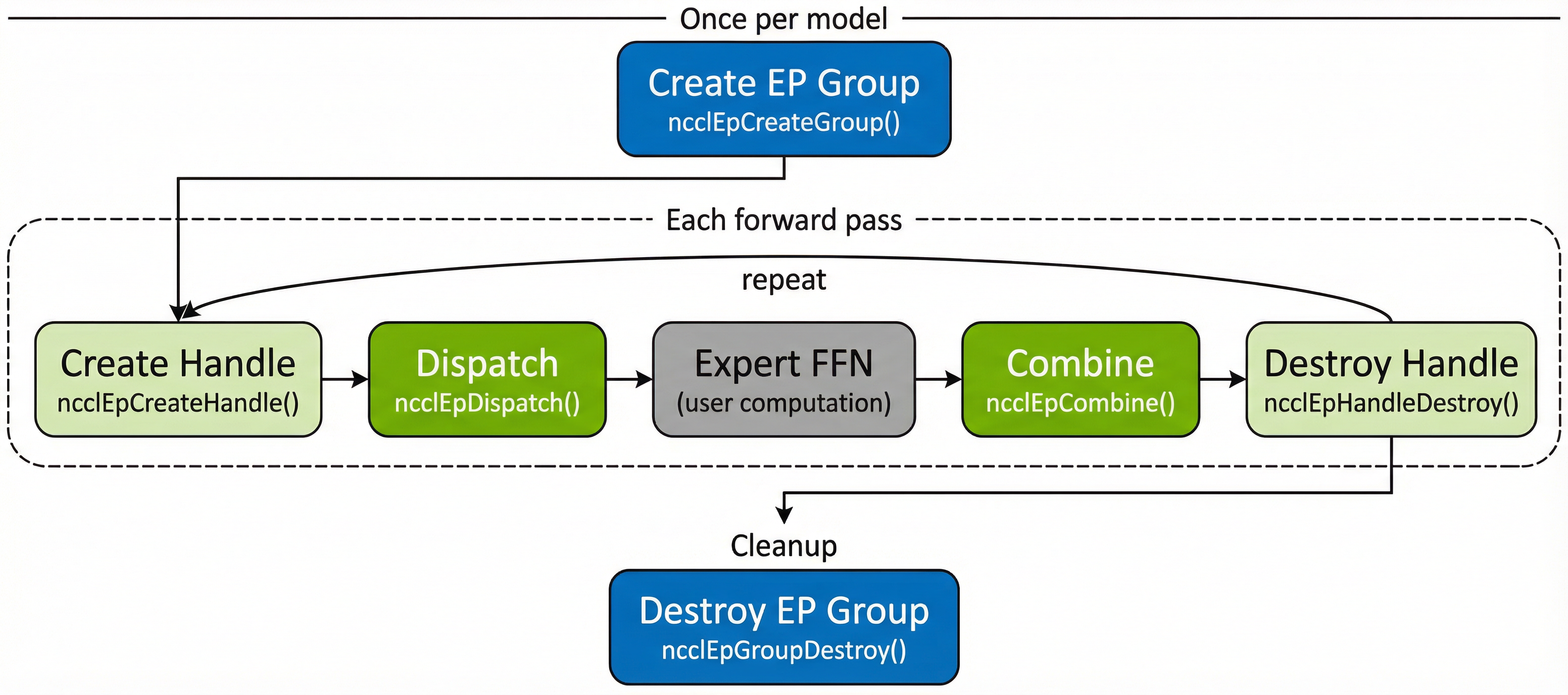}
\caption{NCCL EP execution flow: Group creation (once), followed by 
repeated Handle$\rightarrow$Dispatch$\rightarrow$Expert~FFN$\rightarrow$Combine 
cycles. Staged mode splits dispatch and combine into send and receive 
phases for overlap.}
\label{fig:nccl-ep-flow}
\end{figure}

Table~\ref{tab:nccl-ep-api} summarizes the NCCL EP API, organized by
functional category. NCCL EP provides four core operations that implement
the MoE communication pattern; Figure~\ref{fig:nccl-ep-flow} illustrates
the typical execution flow.

\begin{table*}[t]
\footnotesize
\centering
\caption{NCCL EP C API: Resource management functions create and destroy groups (long-lived) and handles (per-forward-pass). Communication functions perform dispatch and combine operations with optional staged execution for overlap.}
\label{tab:nccl-ep-api}
\begin{tabular}{|l|p{7.5cm}|p{6.0cm}|}
\hline
\textbf{Category} & \textbf{Function} & \textbf{Description} \\
\hline\hline
\multirow{4}{*}{\textit{Resource Management}} 
& \code{ncclEpCreateGroup(group, comm, config, stream, alloc\_fn, free\_fn)} 
& Create MoE group from NCCL communicator; configures algorithm mode (LL/HT), expert count, buffer sizes. Collective call. \\
\cline{2-3}
& \code{ncclEpGroupDestroy(group, stream)} 
& Release group resources including buffers and network connections. \\
\cline{2-3}
& \code{ncclEpCreateHandle(handle, group, topk\_idx, ...)} 
& Create per-forward-pass handle with routing info. \\
\cline{2-3}
& \code{ncclEpHandleDestroy(handle)} 
& Release handle after dispatch/combine sequence completes. \\
\hline
\multirow{3}{*}{\textit{Communication}} 
& \code{ncclEpDispatch(handle, inputs, outputs, send\_only, ...)} 
& Send tokens to assigned experts. \code{send\_only=1} enables staged execution. \\
\cline{2-3}
& \code{ncclEpCombine(handle, inputs, outputs, send\_only, ...)} 
& Gather expert outputs and reduce to original token order. \\
\cline{2-3}
& \code{ncclEpComplete(handle, config, stream)} 
& Finalize staged operation (LL mode). Blocks until receives complete. \\
\hline
\textit{Query} 
& \code{ncclEpHandleGetNumRecvTokens(handle, \&num)} 
& Query received token count (HT mode) for buffer allocation. \\
\hline
\end{tabular}
\end{table*}

\noindent\textbf{Dispatch} (\code{ncclEpDispatch}) sends tokens to their 
assigned experts. It accepts three tensor lists: \emph{inputs} (tokens and 
metadata to route), \emph{outputs} (pre-allocated receive buffers), and 
\emph{local tensors} (optional per-rank metadata such as token counts). 
The \code{send\_only} flag enables staged execution, where dispatch 
initiates data transfers and releases GPU resources without waiting
for the operation to complete, allowing computation-communication overlap.

\noindent\textbf{Combine} (\code{ncclEpCombine}) gathers expert outputs and 
reduces them back to the original token order. It uses the routing state 
stored in the handle to reverse the dispatch permutation. When top-k weights 
are provided via local tensors, combine applies weighted reduction before 
returning tokens to source ranks.
Similarly to dispatch, combine supports staged execution via the \code{send\_only}
flag.

\noindent\textbf{Complete} (\code{ncclEpComplete}) finalizes staged 
operations, waiting for all incoming data transfers initiated by a preceding 
operation with \code{send\_only=1}.

\noindent\textbf{Query} (\code{ncclEpHandleGetNumRecvTokens}) returns the 
actual number of tokens to be received in HT mode, enabling precise output 
buffer allocation when \code{max\_tokens\_per\_rank} is set to auto.

\subsection{Resource Management}
\label{sec:nccl-ep:resources}

NCCL EP employs a two-tier resource hierarchy that separates long-lived 
configuration from per-operation state, enabling efficient resource reuse 
across training iterations or inference requests.

\subsubsection{MoE Group}

The \code{ncclEpGroup\_t} represents the top-level resource container, 
created via \code{ncclEpCreateGroup()} from an existing NCCL communicator.
The group configuration is determined by the algorithm mode (LL or HT),
number of experts, maximum tokens per rank, and buffer-sizing parameters. The group 
encapsulates:
\begin{itemize}
    \item \textbf{Communication algorithm}: The selected mode and
        associated parameters.
    \item \textbf{Communication buffers}: Pre-allocated send/receive 
        buffers sized according to the configuration.
    \item \textbf{Network resources}: Connections (i.e., Queue pairs (QPs))
        to  all peers in the communicator.
\end{itemize}

Group creation is a \emph{collective operation}: all ranks must call 
\code{ncclEpCreateGroup} with compatible configurations. The group 
typically persists for the model's lifetime, created during initialization 
and destroyed at shutdown.

The optional allocator callbacks (\code{alloc\_fn}, \code{free\_fn}) 
enable integration with custom memory pools. When provided, NCCL EP uses 
these callbacks for all internal allocations, allowing systems to 
manage memory across multiple models.

\subsubsection{MoE Handle}

The \code{ncclEpHandle\_t} maintains per-pass state, created via 
\code{ncclEpCreateHandle()} with the routing information (\code{topk\_idx}) 
that specifies which experts each token should be dispatched to. In 
HT mode, handle creation triggers metadata exchange across 
ranks to compute token distribution; in LL mode, this exchange 
occurs implicitly during dispatch.

The handle is shared between matching dispatch and combine operations
of the forward and (for training) backward passes.
This stateful design enables the efficient reuse of the routing
information. Handles are typically created before each forward pass 
and retained as long as the routing remains unchanged.

\subsection{Algorithm Modes}
\label{sec:nccl-ep:modes}

NCCL EP supports two algorithm modes optimized for different workloads. 
The mode is determined at group creation,
either explicitly (via \code{ncclEpGroupConfig\_t.algorithm}) or
(in the future) automatically.

\vspace{0.5em}
\noindent\textbf{LL Mode} is optimized for low latency.
It takes token data as input and produces a three-dimensional,
expert-major output tensor ready to be consumed by subsequent
expert-defined GEMM operations (Figure~\ref{fig:tensor-layout-ll}).
LL execution can be staged (\code{send\_only=1})
enabling expert computation to overlap with data transfer.
Additionally, LL supports overlapping consecutive dispatch
and combine phases through double buffering.

\vspace{0.5em}
\noindent\textbf{HT Mode} is optimized for high throughput and
takes advantage of aggregation and network hierarchy.
For precise buffer sizing, the metadata exchange is initiated across
all ranks during Handle creation.
The 2D output format (Figure~\ref{fig:tensor-layout-ht})
with per-expert token counts ensures deterministic ordering
for reproducible training.

\begin{table}[t]
\footnotesize
\centering
\caption{Comparison of HT and LL kernel modes}
\label{tab:ht-ll-comparison}
\begin{tabular}{@{}lll@{}}
\toprule
\textbf{Aspect} & \textbf{Low-Latency} & \textbf{High-Throughput} \\
\midrule
Target workload & Decode (1--128 tokens) & Train/Prefill (4096+) \\
Communication & All-to-all mesh & Hierarchical NVLink+RDMA \\
Output layout & 3D expert-major & 2D with token counts \\
Reduction & Per-token at receiver & Hierarchical \\
\bottomrule
\end{tabular}
\end{table}

\subsection{Tensor Abstraction}
\label{sec:nccl-ep:tensors}

Traditional NCCL collectives operate on linear buffers with explicit counts 
and data types. MoE operations require richer tensor semantics to describe 
multi-dimensional layouts and distinguish between different tensor roles.

\subsubsection{N-Dimensional Tensor Descriptor}

NCCL EP introduces \code{ncclNDTensor\_t}, a tensor descriptor that 
encodes layout, data type, and semantic role. It captures 
N-dimensional shape and strides (supporting non-contiguous tensors), 
NCCL data type (FP32, BF16, FP16, FP8), a semantic tag indicating the 
tensor's role, and a pointer to GPU memory. This design allows NCCL EP 
to validate tensor shapes, apply mode-specific transformations, and 
handle type conversions automatically.

\subsubsection{Tensor Tags}

Tensor tags are assigned at creation time via \texttt{ncclEpTensorCreate} and 
identify each tensor's role when passed to dispatch and combine operations. 
See Table~\ref{tab:tensor-tags} for the complete set of supported tags.

\begin{table}[h]
\centering
\caption{Tensor tags identifying the role of each \texttt{ncclNDTensor\_t} in \texttt{ncclEpTensorCreate}, \texttt{ncclEpDispatch}, and \texttt{ncclEpCombine}.}
\label{tab:tensor-tags}
\scriptsize
\begin{tabular}{|l|}
\hline
\textcolor{green!50!black}{// token data (input or output)} \\
\texttt{NCCL\_EP\_TENSOR\_TAG\_TOKENS} \\
\hline
\textcolor{green!50!black}{// top-k expert indices} \\
\texttt{NCCL\_EP\_TENSOR\_TAG\_TOPK\_IDX} \\
\hline
\textcolor{green!50!black}{// top-k router weights} \\
\texttt{NCCL\_EP\_TENSOR\_TAG\_TOPK\_WEIGHTS} \\
\hline
\textcolor{green!50!black}{// FP8 quantization scales} \\
\texttt{NCCL\_EP\_TENSOR\_TAG\_SCALES} \\
\hline
\textcolor{green!50!black}{// per-expert token counts (device)} \\
\texttt{NCCL\_EP\_TENSOR\_TAG\_RECV\_EXPERT\_COUNTER\_DEVICE} \\
\hline
\textcolor{green!50!black}{// per-expert token counts (host)} \\
\texttt{NCCL\_EP\_TENSOR\_TAG\_RECV\_EXPERT\_COUNTER\_HOST} \\
\hline
\textcolor{green!50!black}{// default/unused} \\
\texttt{NCCL\_EP\_TENSOR\_TAG\_NONE} \\
\hline
\textcolor{green!50!black}{// per-expert token counts} \\
\texttt{NCCL\_EP\_TENSOR\_TAG\_TOKENS\_PER\_EXPERTS} \\
\hline
\end{tabular}
\end{table}

Tags serve multiple purposes: 
they identify input versus output tensors, 
distinguish between token data and metadata,
and enable NCCL EP to validate tensor shapes.
For example, when dispatching FP8 tokens, the library 
expects a corresponding scales tensor tagged appropriately.
%

\begin{figure}[!htb]
\centering
\includegraphics[width=0.55\columnwidth]{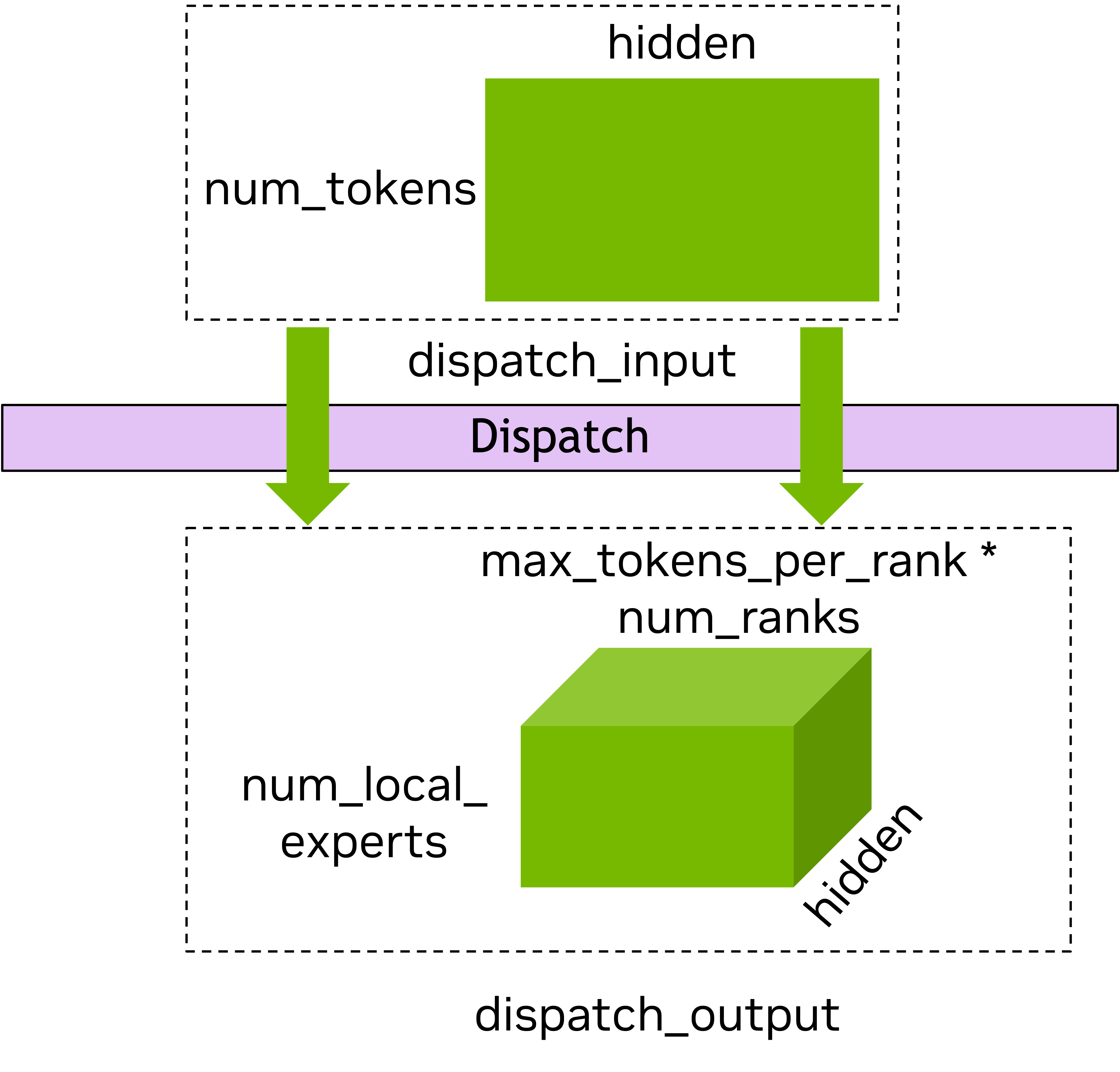}
\caption{LL mode: 2D input to 3D expert-major output.}
\label{fig:tensor-layout-ll}
\end{figure}

\begin{figure} [!htb]
\centering
\includegraphics[width=0.65\columnwidth]{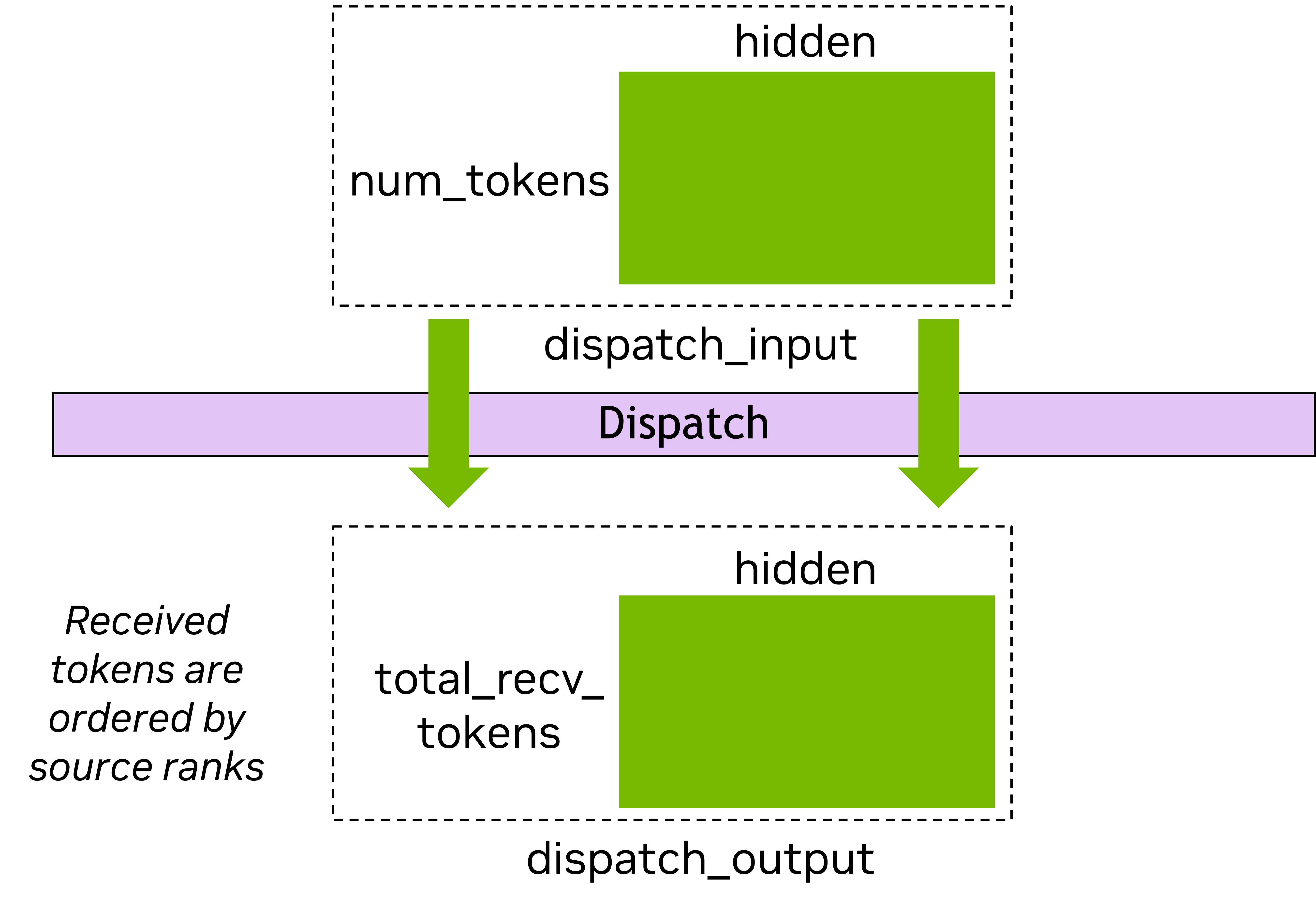}
\caption{HT mode: 2D input to 2D concatenated output.}
\label{fig:tensor-layout-ht}
\end{figure}

\subsubsection{Tensor Layouts}

Input and output tensor layouts differ between algorithm modes 
(Figures~\ref{fig:tensor-layout-ll} and~\ref{fig:tensor-layout-ht}):

\noindent\textbf{Dispatch inputs}: Always 2D with shape 
\code{[num\_tokens $\times$ hidden]} for tokens and 
\code{[num\_tokens $\times$ top\_k]} for routing metadata. All input 
tensors share the same dimension.

\noindent\textbf{Dispatch outputs}:
\begin{itemize}
    \item \textbf{LL mode} (Figure~\ref{fig:tensor-layout-ll}): 3D with shape 
        \code{[num\_local\_experts $\times$ (max\_tokens\_per\_expert * num\_ranks) $\times$ hidden]}, 
        where tokens are groupped by receiving expert. This expert-major 
        layout enables direct input to grouped GEMM operations.
    \item \textbf{HT mode} (Figure~\ref{fig:tensor-layout-ht}): 2D with shape 
        \code{[total\_recv\_tokens $\times$ hidden]}, where tokens from 
        all experts are concatenated. A separate per-expert token count 
        tensor indicates boundaries.
\end{itemize}

\noindent\textbf{Combine}: Input shapes match the corresponding dispatch 
outputs (3D for LL mode as in Figure~\ref{fig:tensor-layout-ll}, 2D for HT 
mode as in Figure~\ref{fig:tensor-layout-ht}); output shapes match dispatch 
inputs, restoring tokens to their original order.

\subsection{Usage Patterns: LL and HT in Training and Inference}
\label{sec:nccl-ep:patterns}

\subsubsection{Training with HT Mode}

MoE handles are shared between forward and backward passes to 
maintain routing state:

%





\noindent
\begin{minipage}{\columnwidth}
\begin{lstlisting}[language=C,basicstyle=\footnotesize\ttfamily]
// Once per model
ncclEpCreateGroup(&group, comm, &config, stream);
// Forward pass: creating one handle/layer in ep[]
for (int l = 0; l < L; ++l) {
    topk = route(tokens);
    ncclEpCreateHandle(&ep[l], group, topk);

    ncclEpDispatch(ep[l], tokens, exp_in, ...);
    expert_ffn_forward(exp_in, exp_out);
    ncclEpCombine(ep[l], exp_out, tok_upd, ...);
    tokens = normalize(tokens, tok_upd);
}
// Backward pass: reusing handles from ep[]
loss = calc_loss(ideal, tokens);
for (int l = L - 1; l >= 0; --l) {
    ncclEpDispatch(ep[l], loss, exp_in, ...);
    expert_ffn_backward(exp_in, exp_out);
    ncclEpCombine(ep[l], exp_out, loss_upd, ...);
    loss = update_loss(loss, loss_upd);

    ncclEpHandleDestroy(ep[l]);
}

\end{lstlisting}
\end{minipage}

\begin{figure}[t]
\centering
\includegraphics[width=\columnwidth]{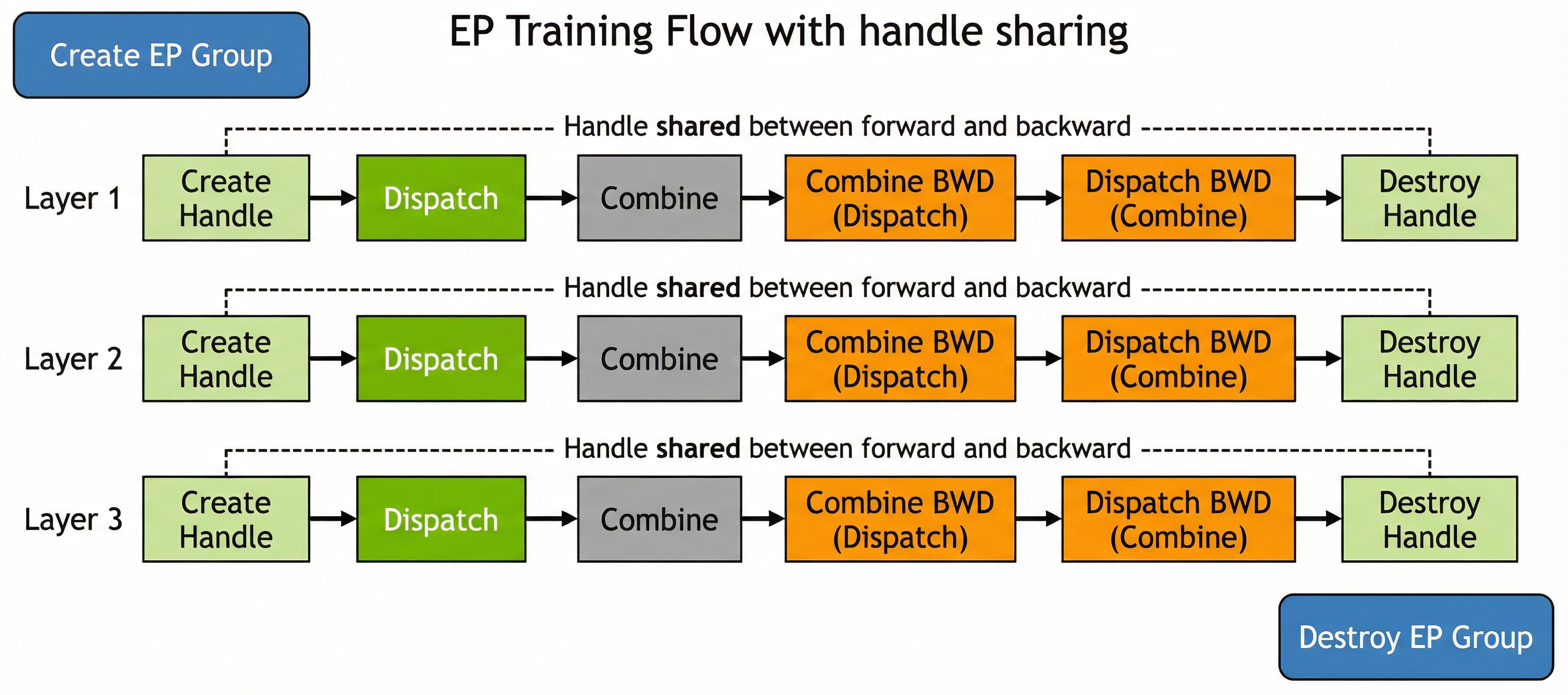}
\caption{MoE training flow: handles are created before dispatch and shared 
between forward and backward passes to maintain routing state.}
\label{fig:training-flow}
\end{figure}

\noindent Figure~\ref{fig:training-flow} illustrates this pattern across 
multiple MoE layers, showing how handles persist through the complete 
forward-backward cycle before destruction.

\subsubsection{Inference with LL Mode}

For decoding, LL mode with staged execution enables overlapping 
multiple micro-batches:

\noindent
\begin{minipage}{\columnwidth}
\begin{lstlisting}[language=C,basicstyle=\footnotesize\ttfamily]
// Micro-batch 0: dispatch (send only)
ncclEpDispatch(ep[0], tokens[0], exp_in[0], ...,
               /*send_only=*/1, stream);

// Micro-batch 1: dispatch while 0 transfers
ncclEpDispatch(ep[1], tokens[1], exp_in[1], ...,
               /*send_only=*/1, stream);

// Complete micro-batch 0 receives and run experts
ncclEpComplete(ep[0], NULL, stream);
expert_ffn_forward(exp_in[0], exp_out[0]);
ncclEpCombine(ep[0], exp_out[0]], tok_upd[0], ...);

// Continue pipeline...
\end{lstlisting}
\end{minipage}

This pattern is particularly effective for inference decode, where multiple 
requests can be batched and pipelined.

\subsection{Python API}
\label{sec:nccl-ep:python}

NCCL EP provides Python bindings through a ctypes-based wrapper that 
requires no compilation, enabling rapid prototyping and framework 
integration. The wrapper mirrors the C API: \code{NCCLLibrary} loads the 
shared library, and methods like \code{ncclEpCreateGroup()}, 
\code{ncclEpDispatch()}, and \code{ncclEpCombine()} map directly to their 
C counterparts. The wrapper automatically converts PyTorch tensors to 
\code{ncclNDTensor\_t} descriptors, handling data type mapping and pointer 
extraction. This enables seamless integration with PyTorch-based frameworks 
like vLLM and SGLang while maintaining the performance benefits of the 
native C.

%% file: text/4-ll-kernels.tex
\section{Low-Latency Kernels}
\label{sec:ll-kernels}

NCCL EP LL mode targets the decode stage of inference, where
small batches (typically 1--128 tokens) make end-to-end latency the primary
concern.
DeepEP~\cite{deepep2025} is the state of the art implementation, that
leverages device-initiated communication,
emphasizing dispatch-combine time reduction and computation-communication overlap.
NCCL EP adopts DeepEP design, replacing the the communication backend with NCCL Device API, and
applying memory footprint and communication optimizations.

\subsection{MoE Model}
\label{sec:ll:moe_model}

NCCL EP assumes an block-wise distribution of $E$ experts 
by $N$ application ranks, with $L = \lceil E / N \rceil$ experts per rank.
On each rank $r$, the attention sub-layer produces a batch of
$B$ fixed size tokens.
For a token index $t \in [0, B)$ on rank $r$,
the MoE router defines a vector $R(r,t)$
of $K$ expert indices to which $t$ is routed.

The logical communication occurs between two sides:
data-parallel ranks (superscript $DP$) and experts ($E$).
Every data-parallel rank $r$ communicates with all experts
forming $E$ valid expert-rank pairs 
$\Gamma^{DP}_r=\{(e, r)|e\in[0, E)\}$,
where $rem^{DP}(e,r) = \lfloor e / L \rfloor$ is the remote rank.
Each expert $e$ communicates with all ranks.
Thus, rank $r$ that hosts $L$ experts also forms $N \cdot L = E$
valid expert-rank pairs
$\Gamma^E_{r}=\{(e,r')|r'\in[0, N), \lfloor e/ N \rfloor = r \}$,
with remote peer defined as $rem^E(e,r) = r$.
The number of tokens communicated between
pair $(e, r)$ is denoted as $m^{s}(e, r)$, 
$s~\in~\{ R, E \}$.

\subsection{DeepEP Communication Model}
\label{sec:ll:communication}

\paragraph{Topology}
LL kernels employ full $N$-to-$N$ mesh connectivity. Every GPU (represented by a rank)
directly communicates with any other GPU through the most efficient available channel,
enabling the high routing flexibility required for dynamic expert distribution in MoE models.

\paragraph{Communication buffer} \label{sec:ll:communication:buffer}
LL mode allocates two internal buffers
used in double-buffering scheme to overlap dispatch and combine operations.
Each buffer is organized into three regions:
(a)~\emph{coordination}--per-$(e,r)$ pair signals, 
(b)~\emph{send}--a bounce buffer to assemble message \emph{payload}, and
(c)~\emph{receive}--for messages from remote ranks.

To avoid synchronization overhead,
each valid expert-rank pair has a dedicated
sub-region in the \emph{coordination} and \emph{receive} regions.
The mapping of a valid pair $(e,r)$ is side-specific. 
For data-parallel rank $r$, it is a source rank-based offset
inside a local expert area: $idx^{DP}(e,r) = (e \bmod N) \cdot N + r$.
On expert, the index is equal to expert:
$idx^E(e,r) = e$.

Each sub-region is divided into $B$ \emph{payload slots}.
The slot size $P$ is determined by the operation's per-token payload message.
The LL dispatch \emph{payload} includes
(a)~a header containing the
source token index,
(b)~token data, and
(c)~optional FP8 quantization scales when precision reduction is enabled.
The combine \emph{payload} includes only token data.
For DeepSeek~v3 model, the token data size is 7168~B for FP8 data type
and 14336~B for BF16; FP8 quantization scales contain 56
floats (224~B).

\paragraph{Communication pattern}
The dispatch and combine kernels have a similar structure.
They are split into \emph{send} and \emph{recv} phases,
coordinated via fine-grained counter \emph{update-and-flush} operations.
Dispatch/send and combine/recv are invoked on
data-parallel side, and dispatch/recv and combine/send - on expert side.
We generalize the calling side via $s~\in~\{ DP, E \}$.

\textbf{Send phase.}
In its \emph{send} region, rank $r$ prepares
payload messages for each input token.
Tokens are mapped to a set of expert-rank pairs
in the operation-specific way.
For dispatch, it is defined by the MoE router:
$\{ (e,r) | e \in R(r,t) \}$;
for combine - via the expert $e$ and rank $r$ dimensions
of a token in the input tensor.
Messages are written directly into 
peer's $r'=rem^s(e,r)$ \emph{receive} sub-region $idx^s(e,r)$
and slot $\alpha \in [0,B)$.
On dispatch, $\alpha$ is computed locally as the next unused index for $(e, r)$.
On combine, it is the token source index $t$.

After all $m^s(e, r)$ tokens corresponding to a pair $(e,r)$ were enqueued,
an \emph{update-and-flush} operation on $idx^s(e, r)$'th
counter of rank $r'$ with operation-specific value $V$ is issued.
For dispatch, $V=(m^s(r,e)+1)$, communicating
the number of valid slots in the respective sub-region
(the additive of $1$ allows non-zero updates when $m^s(e, r) = 0$).
For combine, $V=1$ to flush tokens and signal $DP$ side.

In the \textbf{Receive phase},
rank $r$ monitors the counters of 
all valid expert-rank pairs.
Once a counter is updated for $(e, r') \in \Gamma^s_R$,
the data is extracted to the output tensors
from sub-region $idx^s(e, r')$.
In dispatch, the output tensor layout is identical to the 
\emph{receive} region and is defined by the sub-range index
$idx^E(e, r')$. The mapping of the sub-region slot $\alpha$
to the token index $t$ extracted from the message is cached
on EP handle (for subsequent combine/send).
The combine performs a weighted sum of $K$ expert outcomes for token $t$,
that are located in experts sub-regions in slots with token index $t$.

\subsection{DeepEP Parallelization model}
\label{sec:ll:parallelization}

The LL kernels are executed on a set of $S$ SMs,
each SM $i$ is sub-divided on $W = 32$ warps $\{ \omega_{ij}, i \in [0, S), j \in [0, W) \}$.

A common parallelization strategy found in multiple phases
relies on a split based on expert-rank pairs.
Each SM is assigned $E^{SM} = \lfloor E / S \rfloor$ pairs, 
with each pair allocated with a group 
$\epsilon_e$ of $G = \lfloor W / E^{SM} \rfloor $ warps:

\begin{equation} \label{sec:ll:eq:rank_warp}
\epsilon_e = \{ \omega_{ij} | (i,j) : e = (i \cdot E^{SM} + \lfloor j / G \rfloor \cdot G) \}
\end{equation}
\vspace{-10pt}

\paragraph{Dispatch/send} \label{sec:ll:eq:par:disp_send}

The send phase of the dispatch kernel consists of 3 major steps:
(a)~\emph{Count Tokens}--count the number of tokens per expert,
(b)~\emph{Send Tokens}--communicate payload messages to the target ranks,
(c)~\emph{Update Counters}--coordination via counter \emph{update-and-flush} operation.
\emph{Count Tokens} and \emph{Send Tokens} steps are independent
and executed in parallel.
The \emph{Update Counters} step depends on the completion of the previous two.

\textbf{Token Count.}
A set of warps $\sigma = \{ \omega_{i(W-1)}, i~\in~[0, S) \}$
(one per SM) collaborates to compute the numbers $m^{DP}(e, r_l)$
of tokens routed from local rank $r_l$
to each expert $e~\in~[0, E)$,
with experts evenly distributed across $\sigma$.

\textbf{Send Tokens.}
$B$ tokens in the batch are evenly distributed among $S$ SMs.
For each token assigned to SM $i$,
its payload is collectively packed in the \emph{send} region
warps $\delta_i = \{ \omega_{ij} \} \setminus \sigma$.
Next, messages are sent to $K$ experts
according to the routing table.
Each top-K direction is assigned a dedicated warp in $\delta_i$.

\textbf{Update Counters.}
Finally, the coordination with each remote expert $e \in [0, E)$ is performed.
For a pair $(e, r_{l})$ a warp group $\epsilon_e$ 
(\ref{sec:ll:eq:rank_warp}) is allocated.
It waits for \emph{Count Tokens} and \emph{Send Tokens} steps
for expert $e$ to complete and issues an \emph{update-and-flush}
for $(e, r_l)$.

\paragraph{Dispatch/recv and Combine/send}

These steps are symmetric and utilize the same parallelization strategy,
distributing the work based on expert-rank pairs.
Each pair $(e, r) \in \Gamma^E_r$,
is assigned a warp group
$\epsilon' = \epsilon_{idx^E(e, r)}$ (\ref{sec:ll:eq:rank_warp}).

In dispatch, a warp atomically reserves a contiguous set
of $m^E(e, r)$ slots in the output tensor
corresponding to expert $e$.
The reservation is cached in the EP handle along with token source indices.
This information is used during combine to identify tokens in the input tensor.

Unpacking of tokens from the \emph{receive} region during dispatch,
and packing them to the \emph{send} region during combine
are performed cooperatively by all threads of $\epsilon'$.
The combine payload is homogeneous, allowing the
use of NVIDIA Tensor Memory Accelerator (TMA)
for data transfers.

\paragraph{Combine/recv}
At this step, available warps are split into
reduction groups $\rho$, each containing $R$ warps;
each SM hosts $G'$ groups (typically $G'=2$):

\begin{equation}
\rho = \{ \rho_{m} = \{ w_{ij} : \lfloor j / R \rfloor < G', m = i \cdot G' + \lfloor j / R \rfloor \} \}
\end{equation}
\vspace{-10pt}

\noindent
The set of all received tokens is evenly distributed across $\rho$.
Within each group $\rho_m$, a single warp $ \omega' \in \rho_m$
uses TMA to load chunks of $K$ expert responses into the SM's shared memory
and notifies the remaining warps $\rho' = \rho_m \setminus \omega'$.
Warps of $\rho'$ cooperatively perform weighted reduction of $K$ chunks
into a single chunk, storing it in the output tensor using TMA.
The data flow between $\omega'$ and $\rho'$ is organized as a pipeline.

\subsection{NCCL EP implementation}
\label{sec:ll:ncclep}

NCCL EP implementation is based on DeepEP design presented above,
adopting it for the NCCL infrastructure.
It also uses an optimized communication buffer layout that allows to
significantly reduce memory footprint.

\paragraph{Device-initiated communication}

NCCL EP implements communication channels through
NCCL Device API (Section \ref{sec:background:deviceapi}).
When the destination is in the same NVLink domain (same LSA team),
LL kernels leverage NVLink network using load/store instructions.
Otherwise, the RDMA network is used via NCCL GIN API.

The \emph{update-and-flush} operation is based on
updates of counters associated with expert-rank pairs. 
For NVLink domain, 
a counter for $(e,r)$ pair is implemented via store-release/load-acquire
operations on $idx^s(e, r)$'th integer in the \emph{coordination} region.
For RDMA, a zero-byte NCCL GIN \emph{put}
operation with increment (\code{ncclGin\_SignalAdd}) modifier  is used with 
$\code{signalId} = idx^s(e, r)$, $s \in \{ DP, E \}$.

\paragraph{Communication buffer layout optimization}

In the DeepEP (Section \ref{sec:ll:communication}),
the size of the communication buffer for both dispatch and
combine operations scales as $O(E \cdot B \cdot P)$,
requiring twice the amount for double buffering.
Each attention rank provides up to batch size $B$ tokens,
thus, $B$ slots are reserved for each expert in the \emph{receive} region.
Such design leads to a wasted communication resources, as
all $B$ payload slots are never be occupied for all experts.

To address the issue in the dispatch, we
extend the message header with routing information $R(r,t)$,
avoiding layout-based routing.
In the \textbf{Send Tokens} step (Section \ref{sec:ll:eq:par:disp_send}),
a token is sent once per destination rank, as opposed to once per expert.
The sub-region indexing is unified for dispatch $idx^D(e,r) = r$, not depending
on the calling side.
The dispatch buffer is reduced by a factor of number of local experts $L$,
scaling as $O(N \cdot B \cdot P)$.

The combine benefits from the routing information even further.
For each token $t$ or rank $r$, the index $k$ of the routing entry
that corresponds to expert $e$: $R_k(r,t) = e$ 
is cached during dispatch.
This allows to implement highly efficient combine sub-region indexing 
$idx^C(t,k) = t \cdot K + k$, packing responses with no wasted slots
and scaling as $O(B \cdot K \cdot P)$

The new scheme requires minimal changes in the LL implementation.
As the \emph{coordination} region footprint is low,
the coordination logic was not changed.
The footprint reduction is described by (\ref{sec:ll:eq:mem_red}),
and for $N=64$ ranks, $E=512$ experts, and top-k $K=8$
yields $\approx 14$ times improvement.

\begin{equation} \label{sec:ll:eq:mem_red}
    \frac{2 \cdot E \cdot B \cdot P}{N \cdot B \cdot P + B  \cdot K \cdot P } = 
    \frac{2 \cdot E}{(N + K) }
\end{equation}

\vspace{1em}

%% file: text/5-ht-kernels.tex
\section{High-Throughput Kernels}
\label{sec:ht-kernels}

NCCL EP's HT mode targets training and inference prefilling
workloads where large batch sizes (thousands of tokens) make bandwidth
utilization the primary concern. The HT kernels are adapted from
Hybrid-EP~\cite{hybridep}, which implements efficient hierarchical
communication using TMA for intra-node NVLink transfers and IBGDA for
inter-node RDMA. Our contribution is replacing the IBGDA backend with NCCL GIN,
enabling these kernels to benefit from NCCL's portable networking abstraction
while preserving Hybrid-EP's algorithmic efficiency.

\subsection{Kernel Architecture}

Hybrid-EP uses warp-specialized persistent kernels where each CUDA block
exclusively occupies a single SM and runs a complete data pipeline. Within each
block, different warp groups handle different pipeline stages: loading data
from global memory to shared memory, transferring across the network, and
writing results back. Blocks operate independently on different data chunks
with no inter-block synchronization.

For dispatch, the pipeline consists of three warp groups. One group reads
tokens from local GPU memory into shared memory FIFOs using TMA. Another group
handles inter-node communication, originally using IBGDA to send data to
same-rail GPUs on remote nodes. A third group writes tokens from shared memory
to destination buffers on all GPUs within the local node via NVLink. The
combine operator uses a more complex pipeline with additional warp groups for
hierarchical reduction: partial results are first accumulated within the node,
then transferred across nodes, and finally globally reduced.

This design achieves near-peak bandwidth while using only a small number of
SMs, leaving the majority available for overlapping computation with
communication.

\subsection{NCCL GIN Adaptation}

The key adaptation in NCCL EP replaces Hybrid-EP's IBGDA calls with NCCL GIN
primitives. IBGDA provides direct GPU-initiated RDMA through a low-level
interface that requires explicit NIC topology configuration and memory
registration. NCCL GIN offers similar GPU-initiated networking but through
NCCL's portable abstraction layer, automatically handling topology discovery
and providing consistent semantics across different network fabrics.

The translation is straightforward: IBGDA's RDMA write operations become
\code{ncclGin.put()} calls using window-based addressing. Flow control, which
Hybrid-EP implements through atomic flags in registered memory, translates to
NCCL GIN signal operations for managing head and tail pointers in the circular
buffer protocol. The TMA-based intra-node paths remain unchanged since they
operate independently of the inter-node backend.

One practical benefit of this adaptation is simplified deployment. Hybrid-EP
requires IBGDA to be available and properly configured, with explicit GPU-to-NIC
mappings provided by the user. With NCCL GIN, topology detection is handled by
NCCL, and the kernels work across InfiniBand, RoCE, and other networks that
NCCL supports.

\subsection{Buffer Management}

HT kernels require registered buffers for cross-rank access. For inter-node
communication, memory is registered with NCCL GIN windows; for intra-node,
CUDA IPC handles allow access by other GPUs. Because buffer registration is
expensive, these buffers are allocated once during initialization using
worst-case sizing that assumes all tokens could route to a single rank.

The dispatch output and combine input reside in registered buffers. An
additional device-to-device copy moves data between these registered buffers
and standard PyTorch tensors. In practice, this copy can be fused with the
permute operation required before expert MLPs, hiding the overhead.

\subsection{Comparison with Low-Latency Mode}

Table~\ref{tab:ht-ll-comparison} summarizes the key differences between HT and
LL modes. The HT output is a 2D tensor with shape \code{[num\_recv\_tokens $\times$
hidden]}, along with per-expert token counts indicating boundaries. This
format suits grouped GEMM operations that process variable-sized expert
batches. In contrast, LL mode produces a 3D expert-major layout optimized for
the small batch sizes typical of decode.

%% file: text/6-integration.tex
\section{Framework Integration}
\label{sec:integration}

This section describes NCCL EP integration as a drop-in MoE communication backend
in two production systems: \textbf{Megatron-LM}~\cite{megatronlm} (v0.12) for
training and \textbf{vLLM}~\cite{vllm} (v0.10) for inference. Both frameworks
share a common substrate -- a buffer abstraction over the NCCL EP Python
bindings -- that enables adoption through configuration alone, without
application-level code changes. 
This section serves as an integration guide for adopting NCCL EP in other systems.

\subsection{Common Backend Architecture}
\label{subsec:common-backend}

The integration stack comprises three layers: the NCCL EP C API
(Section~\ref{sec:nccl-ep}), the \code{nccl\_ep} Python package (ctypes bindings
described in Section~\ref{sec:nccl-ep:python}), and a buffer abstraction that adapts
framework tensors to NCCL EP and manages handles, memory, and stream
synchronization. The buffer decouples MoE layers from low-level
NCCL device operations.

\noindent\textbf{Buffer interface.}
The buffer exposes \code{dispatch} and \code{combine} as the primary MoE API,
mapping directly to \code{ncclEpDispatch} and \code{ncclEpCombine}
(Section~\ref{sec:nccl-ep:operations}). \code{NDTensorWrapper} converts framework
tensors to \code{ncclNDTensor\_t} descriptors; communicators come from the
framework's process group. For training backward passes, \emph{cached dispatch}
reuses an existing handle. The buffer also exposes \code{get\_tokens\_per\_expert\_list},
\code{get\_comm\_stream}, and \code{capture} for event synchronization; handles
must be destroyed via \code{destroy\_handle} when no longer needed. Table~\ref{tab:buffer-pseudo}
summarizes the interface.

\begin{table}[t]
\footnotesize
\centering
\caption{Buffer operations: framework-facing API mapping to NCCL EP primitives.}
\label{tab:buffer-pseudo}
\resizebox{\columnwidth}{!}{%
\begin{tabular}{@{}ll@{}}
\toprule
\textbf{Operation} & \textbf{Call signature in Python notation} \\
\midrule
Dispatch & \parbox[t]{0.82\columnwidth}{\texttt{recv\_x, recv\_i, recv\_w, h, ev =}\\\hspace*{1.5em}\texttt{buf.dispatch(x, topk\_i, topk\_w, handle)}} \\
Combine & \parbox[t]{0.82\columnwidth}{\texttt{out, out\_w, ev =}\\\hspace*{1.5em}\texttt{buf.combine(x, h, topk\_w)}} \\
Tokens per expert & \parbox[t]{0.82\columnwidth}{\texttt{tokens\_per\_expert =}\\\hspace*{1.5em}\texttt{buf.get\_tokens\_per\_expert\_list()}} \\
Stream & \parbox[t]{0.82\columnwidth}{\texttt{stream = buf.get\_comm\_stream()}} \\
Capture & \texttt{event = buf.capture()} \\
Destroy & \texttt{buf.destroy\_handle(h)} \\
\bottomrule
\end{tabular}%
}
\end{table}

\noindent\textbf{Memory and streams.}
MoE layers invoke \emph{dispatch} and \emph{combine} dozens of times per forward
pass; naive allocation would trigger frequent CUDA allocations, degrading
performance. The buffer connects NCCL EP's allocator callbacks
(Section~\ref{sec:nccl-ep:resources}) to PyTorch's CUDA caching allocator via
\code{torch.empty(..., device='cuda')}, with pointers stored in a global map for
reuse. Handle destruction releases memory immediately after \emph{combine}
(inference) or after the backward pass (training). A dedicated communication
stream isolates communication from compute, enabling profiling and future overlap.
The current implementation is synchronous: communication waits for compute via
\code{previous\_event}, and compute waits for communication. This ensures
correctness while maintaining infrastructure for asynchronous execution.

\noindent\textbf{Expert counts.}
In HT mode, expert computation requires per-expert token counts computed during
\emph{dispatch}. The buffer allocates pinned, GPU-mapped host memory
(\code{recv\_expert\_counter}) so NCCL EP can write counts visible after stream
synchronization; \code{get\_tokens\_per\_expert\_list} retrieves them without
explicit device-to-host copies.

\subsection{Megatron-LM Integration}
\label{subsec:megatron}

\begin{figure}[!htb]
\centering
\includegraphics[width=0.9\columnwidth]{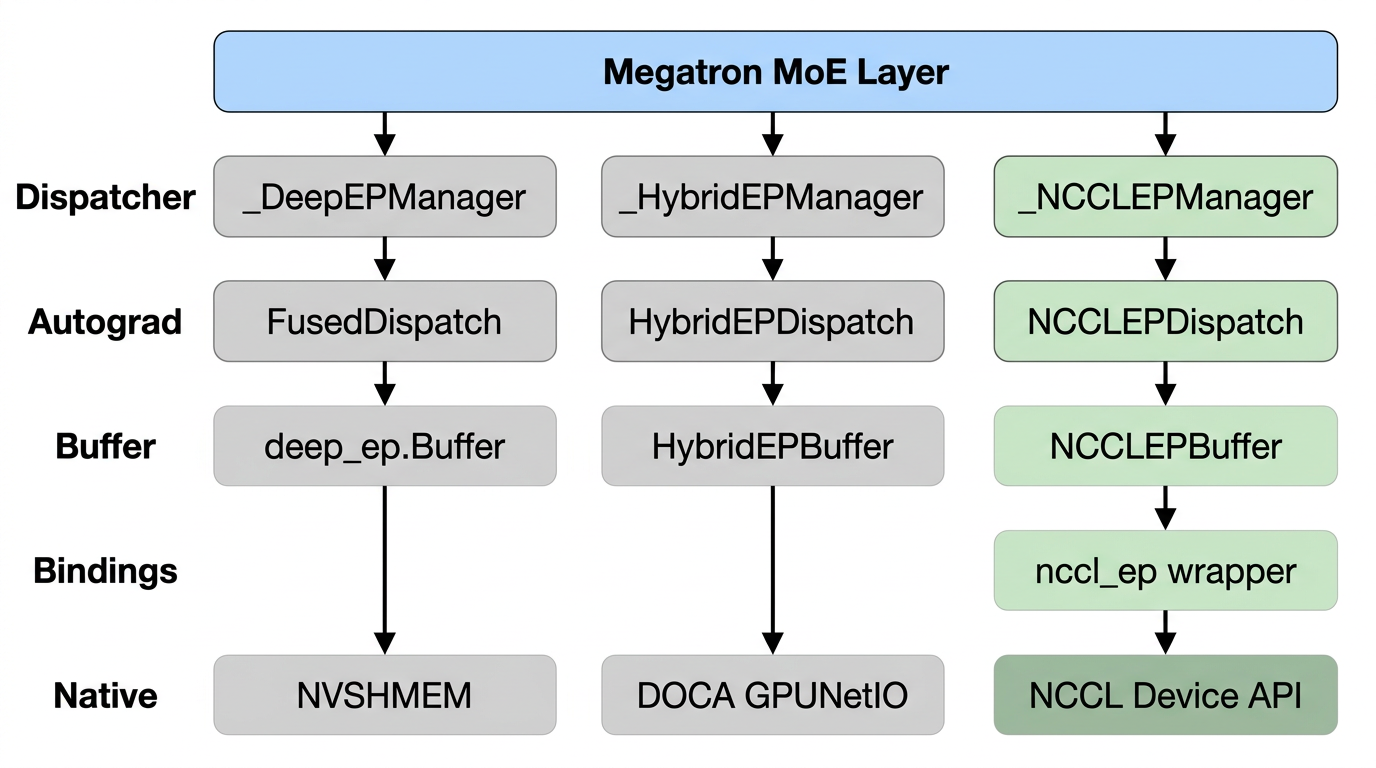}
\caption{NCCL EP in Megatron-LM. The Flex dispatcher supports three
device-initiated backends; NCCL EP (green) is a drop-in alternative to
DeepEP (gray) via compatible APIs at each layer.}
\label{fig:megatron-arch}
\end{figure}

Megatron-LM is a framework used for LLM training.
It offers three MoE communication strategies: AllGather, AllToAll
(standard NCCL collectives), and Flex (device-initiated backends). 
Flex has a plugin architecture and supports DeepEP and Hybrid-EP.
We extend Flex with NCCL EP support,
integrating HT mode in its architecture as shown on Figure~\ref{fig:megatron-arch}.

\textbf{Dispatcher Manager.}
\code{\_NCCLEPManager} adapts Megatron's MoE layer to NCCL EP. It converts Flex's
\emph{multi-hot} routing (sparse $[\mathit{num\_tokens}, \mathit{num\_experts}]$)
to NCCL EP's compact \emph{top-k} format ($[\mathit{num\_tokens}, k]$), casts
routing probabilities to FP32, and otherwise follows the Flex interface for drop-in compatibility.

\textbf{Autograd.}
\code{NCCLEPDispatch} and \code{NCCLEPCombine} wrap buffer operations for
automatic differentiation (autograd). In the forward pass, \emph{dispatch} creates a handle
and redistributes tokens; \emph{combine} aggregates outputs. In the backward
pass, \emph{cached dispatch} redistributes gradients to experts and \emph{combine}
aggregates them; the handle is destroyed after the latter. Handle destruction is
deferred until communication completes to avoid blocking. The buffer records a
CUDA event, queues the handle, and defers destruction until the event signals.

\noindent\textbf{Buffer.}
\code{NCCLEPBuffer} implements the common interface with API compatibility to
DeepEP's buffer. A global singleton stores metadata (process group, expert count,
hidden dimension, \code{max\_tokens\_per\_rank}) for shape inference during
cached operations.

\subsection{vLLM Integration}
\label{subsec:vllm}

vLLM implements expert-parallel MoE inference with pluggable All2All backends
(naive, pplx, DeepEP's HT/LL). NCCL EP integrates as an additional backend
(\code{nccl\_high\_throughput} or \code{nccl\_low\_latency}).
Unlike Megatron, vLLM supports both algorithm modes;
tensor layouts follow the mode-specific formats described in
Section~\ref{sec:nccl-ep:tensors}.

\noindent\textbf{All2All Manager.}
\code{NCCLAll2AllManager} acts as a factory and cache for buffer instances,
keyed by (process group, expert count, hidden dimension,
\code{max\_tokens\_per\_rank}, algorithm). A single \code{NCCLEPBuffer} class
handles both modes, with the algorithm selected at construction.

\noindent\textbf{FusedMoE Layer.}
The FusedMoE layer orchestrates routing, \emph{dispatch}, expert computation, and
\emph{combine}. It integrates NCCL EP through vLLM's \emph{prepare}/\emph{finalize}
abstraction: the prepare phase obtains a buffer from \code{all2all\_manager},
calls \code{dispatch}, and receives a handle; the finalize phase invokes
\code{combine}. \code{NCCLHTPrepareAndFinalize} and \code{NCCLLLPrepareAndFinalize}
wrap \code{NCCLEPBuffer} for each mode. Handles are double-buffered for pipelined
inference and explicitly released after use.

\noindent\textbf{Buffer and activation formats.}
The buffer provides per-expert token counts (device or host) to guide GEMM and
\emph{combine}, avoiding wasted work on padding. In LL mode, the
batched-per-expert layout enables buffer preallocation for lower latency. In HT
mode, the 2D concatenated layout with per-expert counts supports deterministic
ordering.

%% file: text/7-evaluation.tex
\section{Performance Evaluation}
\label{sec:results}

This section evaluates NCCL EP performance against DeepEP, the widely-adopted 
MoE communication library. We present Low-Latency (LL) kernel results for 
inference-decode (Section~\ref{sec:results:ll}) and vLLM integration benchmarks 
(Section~\ref{sec:results:framework}). High-Throughput (HT) kernel evaluation 
and optimization is ongoing; HT results and Megatron-LM integration benchmarks 
will be released after these optimizations are complete.

All experiments are conducted on NVIDIA EOS, an H100-based cluster.
Table~\ref{tab:specs} summarizes the hardware specifications.

\begin{table}[t]
\centering
\caption{Hardware specifications of the EOS evaluation cluster.}
\label{tab:specs}
\footnotesize
\begin{tabular}{@{}l|c@{}}
\hline
\textbf{Specification} & \textbf{EOS} \\
\hline
\\[-0.5em]
Number of Nodes     & 576 \\
GPU Model           & H100 80GB HBM3 \\
GPUs per Node       & 8 (640 GB total) \\
GPU Memory BW       & 3.35 TB/s \\
NVLink              & 4th Gen, 900 GB/s, 18 links/GPU \\
CPU Model           & Xeon 8480CL \\
CPU Cores           & 112 (56$\times$2) \\
CPU Clock           & 2.0 / 3.8 GHz \\
System Memory       & 2 TB \\
InfiniBand          & 8$\times$400 Gbit/s \\
\hline
\end{tabular}
\end{table}

\subsection{Low-Latency Kernels}
\label{sec:results:ll}

LL kernels optimize for MoE inference-decode with small token batches, using 
full all-to-all RDMA mesh connectivity with per-expert signals and hybrid 
NVLink-RDMA paths. NCCL EP currently only supports the hybrid NVLink-RDMA path. 
We evaluate with 256 experts, hidden dimension 7168, 
128 tokens, top-k 8 routing, and BF16 precision for both dispatch and combine 
operations.

\begin{figure}[t]
    \centering
    \includegraphics[width=\columnwidth]{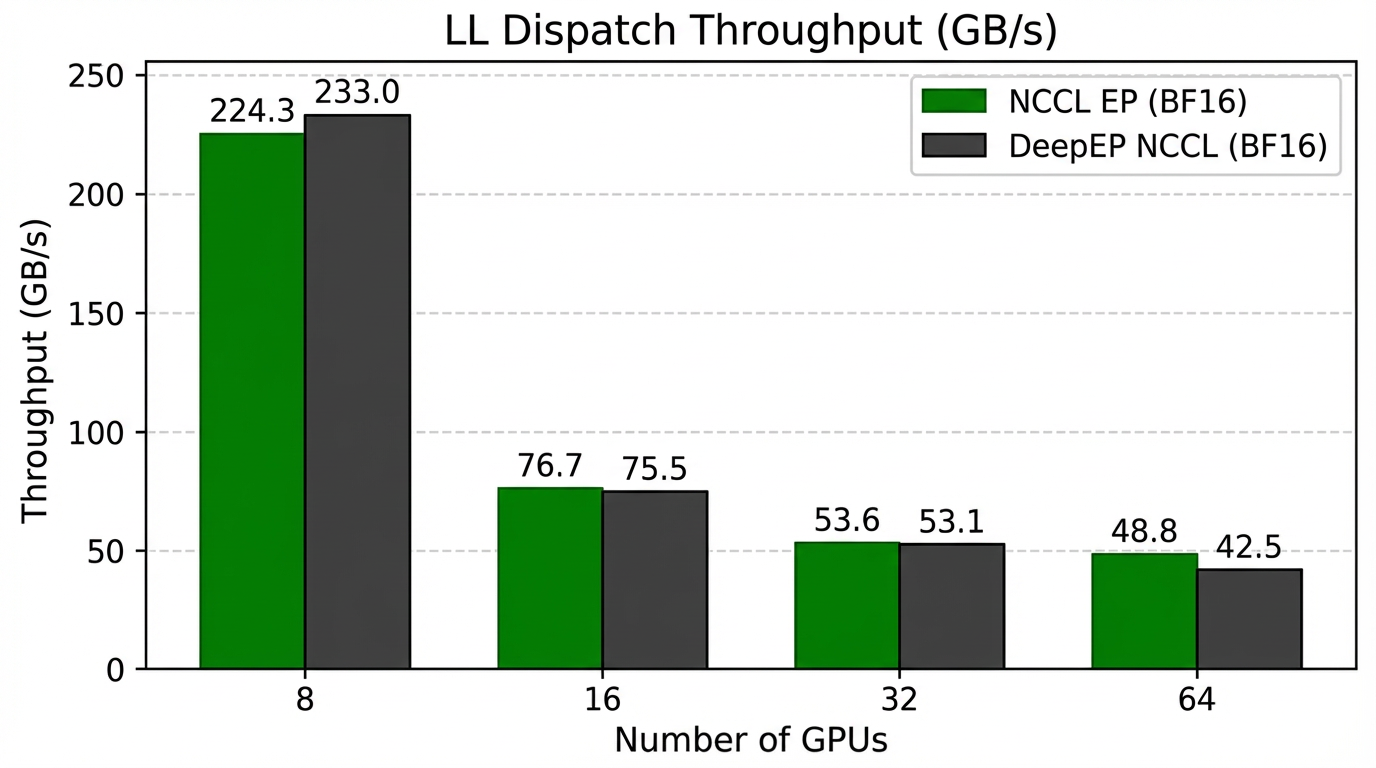}
    \caption{LL dispatch throughput: NCCL EP vs DeepEP.}
    \label{fig:results:ll:dispatch}
\end{figure}

\begin{figure}[t]
    \centering
    \includegraphics[width=\columnwidth]{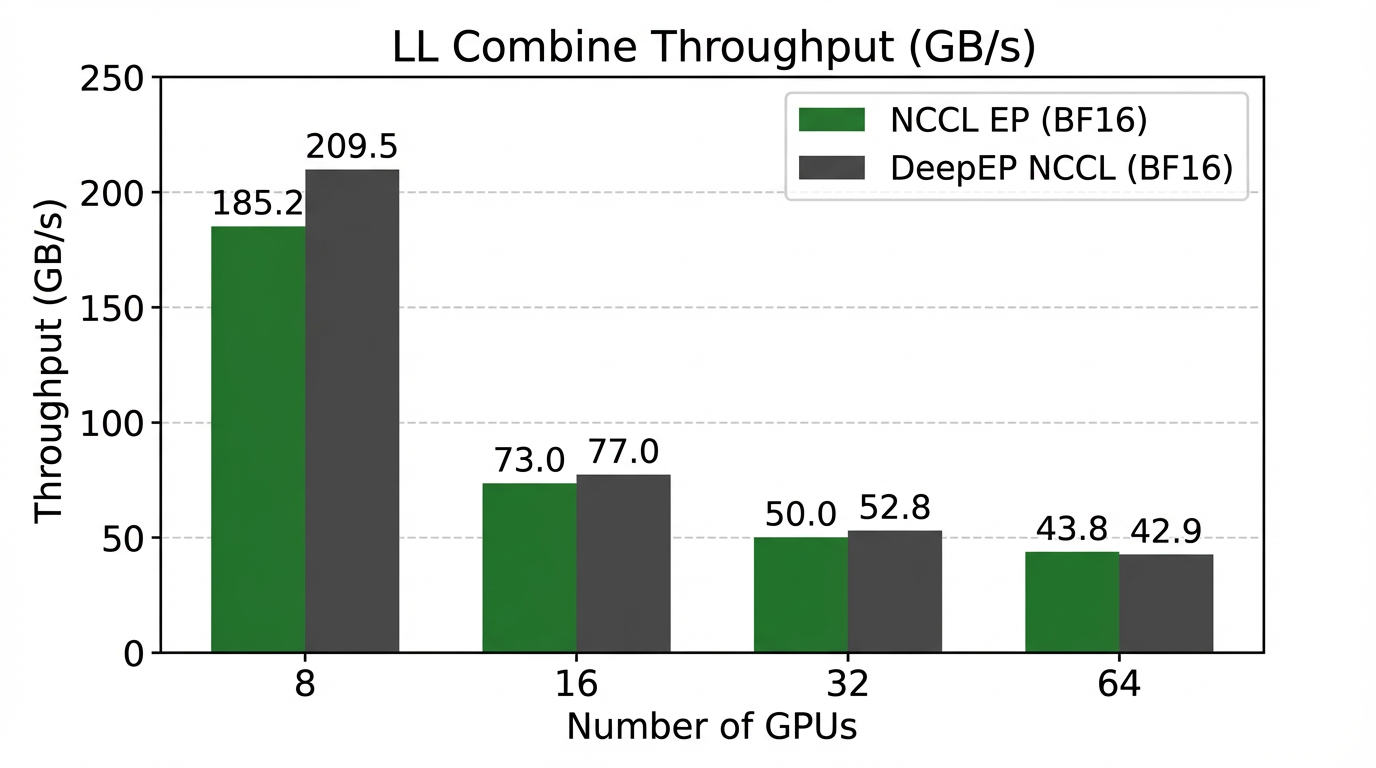}
    \caption{LL combine throughput: NCCL EP vs DeepEP.}
    \label{fig:results:ll:combine}
\end{figure}

Figures~\ref{fig:results:ll:dispatch} and~\ref{fig:results:ll:combine} show 
dispatch and combine throughput comparisons between NCCL EP and DeepEP across 
1 to 8 nodes. In this evaluation, DeepEP is configured to use NCCL (version~2.29) as its 
network backend\footnote{\url{https://github.com/deepseek-ai/DeepEP/pull/521}} 
(rather than NVSHMEM/IBGDA), providing an apples-to-apples comparison 
of the two EP implementations over the same underlying transport~\cite{ncclgin}.
For dispatch, NCCL EP matches or exceeds DeepEP at multi-node scales 
(2--8 nodes), with a notable throughput advantage at 8 nodes (64 GPUs), 
while trailing by $\sim$4\% at single-node.
For combine, NCCL EP trails DeepEP by 5--12\% at 1--4 nodes and 
converges at 8 nodes.

An important methodological note: NCCL EP reports host-side C API time, 
which includes kernel launch overhead, whereas DeepEP reports GPU kernel 
execution time measured via PyTorch's Kineto profiler. Because NCCL EP's 
measurements include this additional overhead, its effective kernel-level 
throughput is higher than the reported values---meaning the dispatch 
advantages are understated and the combine gaps are smaller than shown. 
We chose host-side timing because our benchmark operates at the C layer 
without a dependency on PyTorch. Additionally, at larger scales, 
system-level variance---including network contention, topology-dependent 
routing, and OS scheduling---can introduce up to 5\% variation across runs.

\subsection{vLLM Integration}
\label{sec:results:framework}

Table~\ref{tab:vllm-results} compares end-to-end serving metrics for 
NCCL EP and DeepEP integrated with vLLM on the EOS cluster. Unlike the 
LL kernel evaluation in Section~\ref{sec:results:ll}, which compared both 
implementations over the same NCCL transport, this evaluation runs DeepEP~v1.2.1 
with its default NVSHMEM~3.2.5/IBGDA backend---reflecting the configuration 
users deploy in production. Both backends use vLLM~0.10. We use 1-node (8 GPU), 2-node (16 GPU), and 4-node 
(32 GPU) configurations with the Qwen3-30B-A3B 
model~\cite{qwen3-technical-report}, a 30B-parameter MoE LLM, serving 
1000 requests with a maximum concurrency of~32. All configurations 
use TP=1 with DP=8, 16, and 32 (equal to EP) respectively. Each configuration was 
run 4 times per backend; reported values are means after removing 
statistical outliers via IQR filtering.

\begin{table}[t]
\centering
\caption{vLLM~0.10 serving metrics: NCCL EP vs DeepEP~v1.2.1 with 
NVSHMEM~3.2.5/IBGDA (TP=1, EP=DP=\{8, 16, 32\} (equal to total number of GPUs), 1000 requests, max 
concurrency~32). 
TTFT = time to first token, ITL = inter-token latency, 
TPOT = time per output token.}
\label{tab:vllm-results}
{\footnotesize
\begin{tabular}{@{}llcc@{}}
\toprule
\textbf{Scale} & \textbf{Metric} & \textbf{NCCL EP} & \textbf{DeepEP} \\
\midrule
\multirow{7}{*}{\shortstack[l]{1 node\\(8 GPUs)}}
& Output throughput (tok/s) & 755.1 & 811.2 \\
& Total throughput (tok/s) & 1,509.7 & 1,621.8 \\
& TTFT mean (ms) & 258.5 & 255.6 \\
& TTFT p99 (ms) & 326.5 & 325.5 \\
& ITL mean (ms) & 39.7 & 36.8 \\
& ITL p99 (ms) & 42.6 & 40.6 \\
& TPOT mean (ms) & 39.7 & 36.8 \\
\midrule
\multirow{7}{*}{\shortstack[l]{2 nodes\\(16 GPUs)}}
& Output throughput (tok/s) & 586.1 & 634.0 \\
& Total throughput (tok/s) & 1,171.8 & 1,267.6 \\
& TTFT mean (ms) & 253.9 & 237.4 \\
& TTFT p99 (ms) & 668.7 & 691.5 \\
& ITL mean (ms) & 51.7 & 47.8 \\
& ITL p99 (ms) & 56.6 & 52.8 \\
& TPOT mean (ms) & 51.7 & 47.8 \\
\midrule
\multirow{7}{*}{\shortstack[l]{4 nodes\\(32 GPUs)}}
& Output throughput (tok/s) & 563.3 & 617.0 \\
& Total throughput (tok/s) & 1,126.3 & 1,233.6 \\
& TTFT mean (ms) & 226.5 & 209.5 \\
& TTFT p99 (ms) & 769.1 & 809.9 \\
& ITL mean (ms) & 54.1 & 49.3 \\
& ITL p99 (ms) & 60.3 & 53.6 \\
& TPOT mean (ms) & 54.1 & 49.3 \\
\bottomrule
\end{tabular}}
\end{table}

NCCL EP trails DeepEP by 7--10\% in throughput and 7--9\% in ITL/TPOT 
across all scales. Prior work on NCCL GIN~\cite{ncclgin} showed that 
DeepEP over NCCL GIN and DeepEP over NVSHMEM achieve broadly comparable 
transport-level performance, with each favoring different operations and 
scales. The consistent ITL/TPOT gap is therefore not explained by the 
transport difference; the LL combine overhead observed in 
Section~\ref{sec:results:ll} is a contributing factor. Investigating and 
optimizing this gap is ongoing for future NCCL EP releases.

%% file: text/8-related-work.tex
\section{Related Work}
\label{sec:related}

This section surveys MoE communication libraries and positions NCCL EP
within the landscape. Table~\ref{tab:moe-libraries} in
Section~\ref{sec:background:landscape} summarizes key features; we
elaborate on design choices and trade-offs below.

\subsection{Device-Initiated MoE Libraries}

\noindent\textbf{DeepEP}~\cite{deepep2025} pioneered device-initiated MoE
communication with NVSHMEM and IBGDA. It provides FP8 quantization, 
hook-based overlap, and optimizations for DeepSeek-V3's group-limited 
gating. DeepEP exposes separate Python APIs for LL and HT modes. 
NCCL EP's LL kernels adapt DeepEP's design to NCCL GIN.

\noindent\textbf{Perplexity pplx-kernels}~\cite{pplx-kernels} provides 
device-initiated MoE communication using NVSHMEM with IBGDA, targeting 
low-latency inference. Like DeepEP, it operates outside NCCL and is 
integrated into vLLM as the ``pplx'' backend.

\noindent\textbf{NIXL EP}~\cite{nixl} is NVIDIA's Expert Parallelism module
within the Inference Xfer Library, derived from DeepEP and integrated with
NVIDIA Dynamo for disaggregated inference. It uses its GPUDirect RDMA
backend for zero-copy transfers.

\noindent\textbf{AMD MORI}~\cite{mori} provides ROCm RDMA primitives for
MI-series GPUs and is integrated into SGLang as the ``mori'' backend,
illustrating the need for vendor-specific MoE solutions.

\subsection{Hierarchical and Portable Approaches}

\noindent\textbf{HybridEP}~\cite{hybridep} is NVIDIA's MoE communication
library integrated into Megatron Core v0.15. It uses warp-specialized 
pipelines (G2S, S2G, RDMA stages) with TMA for NVLink and IBGDA for 
inter-node RDMA, requiring only 4--16 SMs to saturate bandwidth. It 
supports DGX Hopper and Grace Blackwell. NCCL EP's HT kernels adapt 
HybridEP's hierarchical design to NCCL GIN.

\noindent\textbf{UCCL-EP}~\cite{uccl-ep} addresses IBGDA portability by
replacing GPU-initiated RDMA with a GPU-CPU control channel: compact
routing commands are sent to CPU proxies that issue GPUDirect RDMA. This
enables deployment on EFA and Broadcom NICs. The proxy approach trades
latency for portability; NCCL EP targets IBGDA-capable hardware with
NCCL ecosystem integration.

\noindent\textbf{Meta NCCLX AllToAllvDynamic}~\cite{ncclx} supports
GPU-resident routing metadata for dynamic token dispatch without CPU
involvement. It powers Meta's services at 100,000+ GPU scale but is a
full NCCL replacement rather than an extension, limiting availability
outside Meta's infrastructure.

\subsection{Inference and Serving Platforms}

\noindent\textbf{Mooncake}~\cite{mooncake} is Moonshot AI's KVCache-centric
platform for Kimi, with a Transfer Engine for RDMA zero-copy transfers.
Its primitives have been extended for MoE dispatch/combine and integrated
into vLLM and SGLang as the ``mooncake'' backend.

\noindent\textbf{Inference frameworks} such as vLLM~\cite{vllm} and
SGLang~\cite{sglang} support multiple AllToAll backends
(\texttt{pplx}, \texttt{deepep}, \texttt{mooncake}, \texttt{mori},
\texttt{flashinfer}). The AllGather/ReduceScatter fallback works for small
EP sizes within NVLink domains but does not scale to multi-node. These
frameworks motivate drop-in MoE backends that integrate with existing
infrastructure.

\subsection{NCCL EP's Positioning}

NCCL EP differs from existing solutions in three key ways:

\begin{enumerate}
    \item \textbf{NCCL Integration}: Unlike standalone libraries
        (DeepEP, pplx-kernels, Mooncake) that operate outside NCCL,
        or proprietary replacements (NCCLX), NCCL EP extends NCCL
        with native MoE primitives via the Device API, thereby
        preserving ecosystem compatibility and unified resource
        management.

    \item \textbf{Unified API}: DeepEP, HybridEP, and pplx-kernels
        require separate interfaces for LL and
        HT modes; NCCL EP offers unified
        \texttt{ncclEpDispatch}/\texttt{ncclEpCombine} primitives
        with the algorithm mode selected at group creation time.

    \item \textbf{C and Python Interfaces}: NCCL EP offers both a
        C API for framework developers and Python bindings for rapid
        prototyping, providing greater flexibility than Python-only
        (DeepEP) or C++-only (HybridEP) alternatives.
\end{enumerate}

By combining proven kernel techniques from DeepEP (LL) and
HybridEP (HT) with NCCL's production infrastructure, NCCL EP
delivers high-performance MoE communication within the standard GPU
communication ecosystem.

%% file: text/9-conclusions.tex
\section{Conclusions and Future Work}
\label{sec:conclusion}

This paper presents NCCL EP, a ground-up MoE communication library built entirely on NCCL infrastructure.
NCCL EP provides unified \texttt{ncclEpDispatch} and
\texttt{ncclEpCombine} primitives with both C and Python interfaces, addressing the ecosystem fragmentation of existing MoE libraries.
Rather than separate communication stacks and mode-specific APIs, it offers a
single interface that supports both low-latency mode (for inference decoding)
and high-throughput mode (for training and prefilling), with the algorithm
selected at group creation.

From the implementation perspective, NCCL EP adopts best industry practices.
For low-latency, an optimized version of DeepEP design is proposed, which demonstrates on-par performance while  requiring an order of magnitude less memory.
For high-throughput scenario, NCCL EP integrates NVIDIA HybridEP solution.
The GPU-initiated communication is leveraging
NCCL Device API that offers 
Load/Store Accessible mode for NVLink communication, and
GPU-Initiated Networking (GIN) for inter-node RDMA.

Our evaluation shows that NCCL EP achieves comparable low-latency kernel
performance to DeepEP over NCCL across multi-node configurations. End-to-end
vLLM integration demonstrates equivalent serving performance, with throughput
and latency on par with DeepEP. These results demonstrate that building MoE
communication natively within NCCL preserves performance while delivering
ecosystem benefits: topology awareness, fault tolerance, and integration with
NCCL's production infrastructure.

NCCL EP is currently an experimental API. Megatron-LM integration is
complete; training performance remains limited by ongoing high-throughput
kernel optimization for multi-node setups. Community engagement and feedback
will be essential to refine the API, identify performance opportunities, and
ensure NCCL EP meets the diverse requirements of MoE workloads across
training and inference deployments.

%% file: text/ack.tex
\section*{Acknowledgments}


The authors acknowledge the use of Cursor AI to assist in the writing 
and editing of this manuscript. The authors reviewed and approved all 
content for accuracy and originality.